\documentclass[aps,prx,twocolumn,superscriptaddress]{revtex4-1}
\usepackage{amsfonts,amsmath,amssymb,graphicx}
\usepackage[qm]{qcircuit}
\usepackage{color}
\usepackage{longtable,array}

\newtheorem{dfn}{Definition}
\newtheorem{thm}{Theorem}

\newtheorem{lem}{Lemma}
\usepackage{mathtools}
\newcommand{\ket}[1]{|#1\rangle}
\newcommand{\bra}[1]{\langle #1|}
\newcommand{\braket}[2]{\langle #1 | #2 \rangle}
\newcommand{\cO}{\mathcal{O}} 
\newcommand{\cB}{\mathcal{B}}
\newcommand{\cC}{\mathcal{C}}
\newcommand{\RR}{\mathbb{R}}
\newcommand{\cD}{\mathcal{D}}
\newcommand{\cU}{\mathcal{U}}
\newcommand{\cE}{\mathcal{E}}
\newcommand{\cc}{\mathfrak c}
\newcommand{\csu}{\mathfrak{su}}
\newcommand{\cM}{\mathcal{M}}
\newcommand{\cP}{\mathcal{P}}
\newcommand{\ba}{\hat{a}}
\newcommand{\bb}{\hat{b}}
\newcommand{\bc}{\hat{c}}
\newcommand{\bd}{\hat{d}}
\newcommand{\cG}{\mathcal G}
\newcommand{\cZ}{\mathcal Z}
\newcommand{\Cp}{\ket{\mathcal{C}_\alpha^+}}
\newcommand{\Cm}{\ket{\mathcal{C}_\alpha^-}}
\newcommand{\Cpd}{\bra{\mathcal{C}_\alpha^+}}
\newcommand{\Cmd}{\bra{\mathcal{C}_\alpha^-}}
\newcommand{\Cone}{\mathcal{C}_1}
\newcommand{\Ctwo}{\mathcal{C}_2}

\usepackage{ulem}

\begin{document}

\title{Repetition Cat Qubits for Fault-Tolerant Quantum Computation}

\author{J\'er\'emie Guillaud}
\email[]{jeremie.guillaud@inria.fr}
\affiliation{QUANTIC Team, Inria Paris, 2 rue Simone Iff, 75012 Paris, France}
\author{Mazyar Mirrahimi}
\affiliation{QUANTIC Team, Inria Paris, 2 rue Simone Iff, 75012 Paris, France}

\date{\today}

\begin{abstract}
We present a 1D repetition code based on the so-called cat qubits as a viable approach toward hardware-efficient universal and fault-tolerant quantum computation. The cat qubits that are stabilized by a two-photon driven-dissipative process, exhibit a tunable noise bias where the effective bit-flip errors are exponentially suppressed with the average number of photons. We propose a realization of a set of gates on the cat qubits that preserve such a noise bias. Combining these base qubit operations, we build, at the level of the repetition cat qubit, a universal set of fully protected logical gates. This set includes single-qubit preparations and measurements, NOT, controlled-NOT, and controlled-controlled-NOT (Toffoli) gates.  Remarkably, this construction avoids the costly magic state  preparation, distillation, and injection. Finally, all required operations on the cat qubits could be performed with slight modifications of existing experimental setups. 
\end{abstract}

\maketitle

\section{Introduction}\label{sec:intro}
Quantum computers are expected to efficiently solve classically intractable problems. The realization of a large-scale quantum computer is challenging because the noise induced by the uncontrolled interactions of the computer's components with the environment destroys the fragile quantum features responsible for the expected speedup. Indeed, all algorithms with theoretically proven quantum speedup require some level of protection against decoherence. The theory of fault-tolerant quantum computation~\cite{Shor,Cambell-Terhal-Nature-2017} precisely addresses this issue. Quantum-error-correcting codes (QECCs) are designed~\cite{Shor1995,Steane1996} such that errors induced by the environment do not affect the quantum information. These codes operate by the ``fight entanglement with entanglement'' mantra: Natural errors arising in physical systems being typically local, the quantum information to be protected is encoded in nonlocal entangled states such that it becomes unlikely that errors can corrupt it, the most popular being the surface code~\cite{Bravyi1998,Dennis2002,Fowler2012}. The crux of the theory of quantum fault tolerance is the \textit{ threshold theorem}: Arbitrarily long quantum computations can be performed reliably provided the noise afflicting the computer's physical components is below a constant value called the \textit{accuracy threshold}~\cite{Aharonov2008, Kitaev1997, Knill1998, Preskill1998, Aliferis2006}.

In theory, QECCs provide, when operated below the threshold, an arbitrarily good protection against the noise, thus solving the decoherence problem. However, their actual implementation comes at the price of tremendous physical resources to reach a sufficient level of protection. This trade-off between the degree of protection provided by a QECC versus the increase in physical components needed for its implementation is the \textit{resource overhead problem}. Realistic  approaches to quantum computation must deal with this issue. In this light, continuous-variable (CV) systems (such as a harmonic oscillator), in which an infinite-dimensional Hilbert space is readily available to protect and process quantum information,  seem to have a head start over discrete-variable (DV) systems that have only a finite-dimensional Hilbert space. There are many different CV encodings, usually involving the superposition of certain specific states of a harmonic oscillator, such as position and momentum eigenstates~\cite{Lloyd1998, Braunstein1998, Gottesman2001}, Fock states~\cite{Chuang1997, Knill2001, Michael2016}, or coherent states~\cite{Cochrane1999, Leghtas2013}. 

The latter encoding, known as cat codes, has been the subject of intensive theoretical and experimental research throughout the past years. A highlight of this research is the first realization of quantum error correction for a quantum memory at the break-even point~\cite{Ofek-Petrenko-Nature-2016}. Some initial theoretical proposals~\cite{Mirrahimi2014,Albert-PRL-2016,Puri-Blais-2017,Cohen-Mirrahimi-PRL-2017,Puri-Girvin-2018} and a few experiments~\cite{Leghtas2015,Touzard-PRX-2018,Rosenblum-Science-2018} indicate that this encoding can be extended to a logical qubit with the possibility of performing protected logical gates. However, the protection   remains  limited to first-order errors due to photon loss, the major decay channel of a superconducting cavity. Two major questions are in order. Can we extend this encoding to a fully fault-tolerant and universal quantum computation protocol? Can we benefit from the advantages of the infinite-dimensional Hilbert space of the harmonic oscillator to achieve a hardware-efficient scaling? This paper aims at answering these questions by  putting forward a new direction toward fault tolerance with a highly economic hardware complexity. 

The concatenation of a CV code, such as the single-mode Gottesman-Kitaev-Preskill code, with a DV one, such as the surface code, has been recently investigated by various groups~\cite{fukui-prl-2017,fukui-prx-2018,vuillot-terhal-2018}. The main idea behind these proposals is that using CV codes as base qubits leads to important improvements in the accuracy threshold of the DV encoding. Note, however, that one can expect this improvement to be less significant in a realistic circuit-based noise model~\cite{Cambell-Terhal-Nature-2017,vuillot-terhal-2018}. Our approach is different: By employing a cat code as the base qubit, the noise structure is modified in such a way that quantum error correction becomes of similar complexity as classical error correction and can be performed using a simple repetition code. Importantly, this specific noise structure can be preserved for a set of fundamental operations which at the level of the repetition code lead to a universal set of protected logical gates.

The pumped (stabilized) cat qubits are known to benefit from a \textit{noise bias}~\cite{Mirrahimi2014,Puri-Blais-2017}. More precisely, one effective error channel (the bit flips for the encoding of this paper) is suppressed exponentially with the ``size'' (the mean number of photons) of the Schr\"odinger cat states.  This suppression is expected to be valid for a large class of physical noise processes with a local effect on the phase space of a harmonic oscillator~\cite{cohen-thesis-2017}. This class includes, but is not limited to, photon loss, thermal excitations, photon dephasing, and various nonlinearities induced by a coupling to a Josephson junction. Recent experiments, in the framework of quantum superconducting circuits, observe such an exponential suppression~\cite{lescanne-dephasing-2019}.

Previous works show that such a noise asymmetry increases the accuracy threshold of various encodings when it is correctly exploited~\cite{Aliferis2008, Tuckett2018}. These theoretical proposals aim at physical qubits (e.g. NV centers in diamonds~\cite{Childress2013}) that naturally benefit from such a noise bias. In the case of pumped cat qubits, the noise bias is tunable and can reach extremely high values.  Even more remarkably, the extra degree of freedom associated to the complex amplitude of the coherent states defining the cat qubit, can be exploited to overcome some no-go theorems (see the Appendix) for bias preserving  operations.  In other words, the infinite-dimensional Hilbert space of the harmonic oscillator that supports the cat qubit state can be exploited to perform various nontrivial gates (such as CNOT and Toffoli) while preserving the  noise bias. This ability was first observed for the CNOT gate in Ref.~\cite{Puri-Girvin-2018} in the case of nondissipative pumped cats. These features lead to a change of paradigm that significantly simplifies the picture toward a universal set of protected logical gates. First, we obtain a universal set of protected logical gates at the level of a simple repetition code. Second, the circuits for implementing the Clifford gates are greatly simplified, and the overhead requirements are significantly reduced. Finally, there is no need for magic state preparation and distillation, even for non-Clifford gates.

In the next section, we overview the basics of encoding quantum information in two-photon pumped cat states, and we recall the reasons behind the exponential suppression of effective bit-flip errors. Next, in Sec.~\ref{sec:overall_scheme}, we present our approach toward hardware-efficient and fault-tolerant quantum computation based on encoding the information in a simple repetition code of cat qubits. We provide a detailed comparison with Ref.~\cite{Aliferis2008} to point out the new capabilities granted by the use of cat qubits as base qubits. In Sec.~\ref{sec:physical_gates}, we explain how to realize key fundamental operations that preserve the noise bias at the level of cat qubits. This set of fundamental operations includes nontrivial operations such as CNOT and Toffoli that are not achievable in a bias-preserving manner for regular qubits with biased noise.  In Sec.~\ref{sec:logical_gates}, we provide the details on how to combine the fundamental operations to obtain a universal set of protected logical gates at the level of the repetition code, thus providing a road map for hardware-efficient fully protected quantum computation. Next, in Sec.~\ref{sec:error_analysis}, we analyze the performance of the fundamental operations in the presence of a realistic noise model and discuss the fault tolerance. In Sec.~\ref{sec:experimental}, we provide a road map toward experimental realization of various components. We argue that all  components or components of similar complexity are already been implemented separately and with efficiencies or fidelities at the level of the accuracy threshold for a realistic circuit-based error model. We conclude in Sec.~\ref{sec:conc} providing  further research directions.

\section{Pumped/stabilized cats as qubits with biased noise}\label{sec:bias}
The proposal in this paper is focused on cat qubits stabilized by two-photon driven dissipation~\cite{Mirrahimi2014}. All concepts can also be adapted to the so-called Kerr cats, where the protection is ensured through a Kerr-type Hamiltonian and two-photon drives~\cite{Puri-Blais-2017}. 

Driving a nonlinear interaction Hamiltonian between a harmonic oscillator and its bath, it is possible to engineer a nonstandard situation where the  oscillator dominantly gains or loses photons in pairs~\cite{Mirrahimi2014,Leghtas2015,Touzard-PRX-2018,lescanne-dephasing-2019}.  The master equation governing the evolution of the oscillator is given by
\begin{equation}
\dot{\rho} = [\epsilon_{2\text{ph}}\hat{a}^{\dag 2} - \epsilon^*_{2\text{ph}}\hat{a}^2, \rho] + \kappa_{2\text{ph}} \mathcal{D}[\hat{a}^2]\rho
\label{eq:2photon}
\end{equation}
with $\mathcal{D}[\hat{L}]\rho = \hat{L} \rho \hat{L}^\dag - \frac12\hat{L}^\dag\hat{L}\rho - \frac12\rho\hat{L}^\dag\hat{L}$. It has been shown~\cite{Gilles1994} that this dynamics stabilizes a two-dimensional Hilbert space spanned by the coherent states $\{ \ket\alpha, \ket{-\alpha}\}$, where $\alpha$ is a complex number fixed by the ratio of the amplitude of the drive to the two-photon dissipation rate: $\alpha = \sqrt{2\epsilon_{2\text{ph}}/\kappa_{2\text{ph}}}$. Equivalently, this manifold is generated by the in-phase and out-of-phase superpositions of these coherent states, known as Schr\"odinger cat states $\Cp:= \mathcal{N}_+ (\ket\alpha + \ket{-\alpha})$, $\Cm:= \mathcal{N}_- (\ket\alpha - \ket{-\alpha})$ where $ \mathcal{N}_\pm := [2(1\pm e^{-2|\alpha|^2})]^{-1/2}$. Expanding the coherent states in the Fock state basis, one can note that the in-phase (respectively, out-of-phase) superposition spans even (respectively, odd) Fock states only; thus, the cat state $\Cp$ (respectively, $\Cm$) is referred to as the even cat (respectively, odd cat). The steady state of Eq.(\ref{eq:2photon}), denoted $\rho_\infty$, can be computed from the initial state $\rho_0$ using the invariants of the dynamics~\cite{Mirrahimi2014}: 
\begin{equation}
\begin{split}
\rho_\infty = c_{++}\Cp\Cpd + c_{--}\Cm\Cmd \\
+ c_{+-}\Cp\Cmd + c^*_{+-}\Cm\Cpd
\end{split}
\label{eq:ss}
\end{equation}
where $c_{++}$, $c_{--}$ and $c_{+-}$ are conserved quantities that are entirely determined by the initial state $\rho_0$.

\begin{figure}[h]
\includegraphics[width=.5\textwidth]{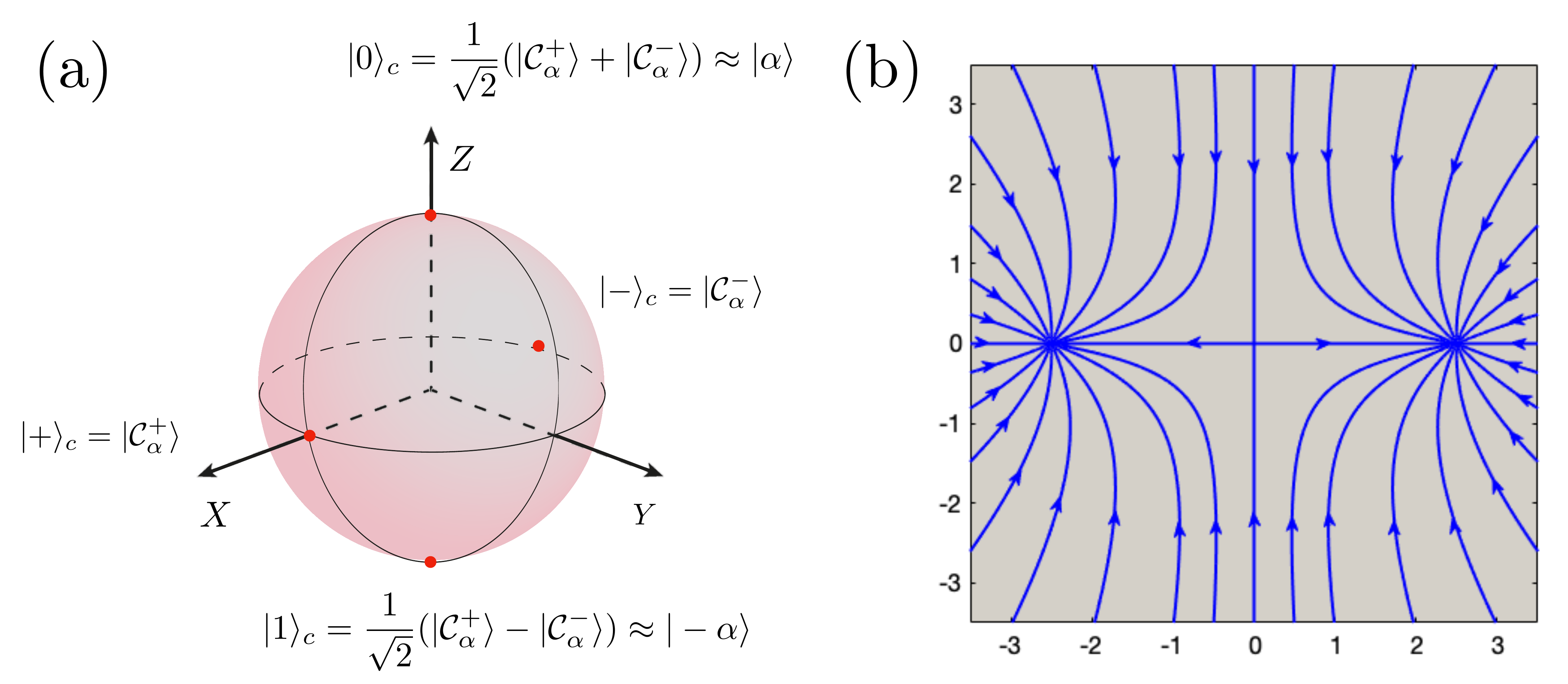}
\caption{(a) Bloch sphere representation of a cat qubit. (b) Vector field associated to the semiclassical dynamics behind the master equation~\eqref{eq:2photon} represented in the phase space of the harmonic oscillator. This vector field governs the dynamics of coherent states. It admits two stable equilibria $\ket{\pm\alpha}$ and one saddle point at zero. The exponential suppression of the bit-flip errors can be understood by the fact that any local perturbation of the state $\ket{0}_c\approx\ket{\alpha}$ (respectively, $\ket{1}_c\approx\ket{-\alpha}$)  keeps the state in the domain of attraction of $\ket{\alpha}$ (respectively, $\ket{-\alpha}$). \label{fig:qubit_bias}}
\end{figure}

The cat qubit states are defined as (see Fig.~\ref{fig:qubit_bias}) $\ket{\pm}_c= \ket{\mathcal{C}_\alpha^\pm}$, or, equivalently as
\begin{align*}
\ket0_c= \tfrac{1}{\sqrt2} (\Cp + \Cm) &= \ket\alpha+\cO[\exp(-2|\alpha|^2)],\\
\ket1_c= \tfrac{1}{\sqrt2} (\Cp - \Cm) &= \ket{-\alpha}+\cO[\exp(-2|\alpha|^2)].
\end{align*}
Note that, with respect to our previous publications (e.g. Ref.~\cite{Mirrahimi2014}), we change the computational basis to the dual basis along the $X$ axis. This choice is motivated by the simplifications in the presentation of the implemented logical gates.

In terms of quantum information processing, the interest of this cat qubit lies in the fact that its physical implementation endows it with a natural protection. As soon as the action of a noise process is local in the phase space of the harmonic oscillator, the effective bit-flip errors (jumps between $\ket{0}_c$ and $\ket{1}_c$) are exponentially suppressed with $2|\alpha|^2$~\cite{Mirrahimi2014,cohen-thesis-2017}. This protection is illustrated in Fig.~\ref{fig:qubit_bias}(b), where the vector field associated to the semiclassical dynamics of a coherent state governed by Eq.~\eqref{eq:2photon} is plotted in the phase space of the  oscillator. Any noise process that perturbs the coherent state $\ket{\pm\alpha}$ locally in the phase space, keeps it in the attraction domain of the departing point $\ket{\pm\alpha}$. Such a protection is similar to the one achieved by topological qubits such as Majorana fermions, but the nonlocality of information in the phase space, is here engineered through the particular driven-dissipative process of the harmonic oscillator. In particular, the nonlocality can be tuned by modifying the cat ``size,'' given by the mean number of photons $|\alpha|^2$. This mean number is itself easily modulated by controlling the strength, $\epsilon_{2\text{ph}}$, of the two-photon drive. The local character of the noise processes is an omnipresent concept in information protection, and, in the case of superconducting oscillators, it includes various  mechanisms such as photon loss, thermal excitations, photon dephasing and nonlinear interaction Hamiltonians induced by Josephson circuits. Indeed, the cosine Hamiltonian of a Josephson junction $\hat H = E_J\cos[\varphi_a(\hat a+\hat a^\dag)]$ represents a bounded operator in the phase space of a mode $\hat a$. In this sense, and over short time steps, it can only lead to a local shift of the state of the harmonic oscillator in the phase space. Furthermore, the rate of diffusion  remains bounded when the cat size increases. The bit flips due to such local shifts are exponentially suppressed in the presence of the two-photon process(see~\cite{cohen-thesis-2017} for more details).

Note, however, that phase flips, or, equivalently, jumps between even-parity cat state $\Cp$ and the odd-parity one $\Cm$, can  be induced by noise mechanisms such as photon loss or thermal excitations.  As a result, an increase of the mean photon number (in order to suppress the bit-flip errors) comes at the expense of higher phase-flip rates. This rate increase is, however, expected to be only linear with respect to $|\alpha|^2$. The noise bias $\exp(-2|\alpha|^2)/|\alpha|^2$ of cat qubits is therefore tunable with the cat size. Some experimental indications of such a tunable bias have been recently observed~\cite{lescanne-dephasing-2019}.

This protection can also be achieved through a nondissipative process using a strong Kerr-type nonlinearity and  two-photon driving~\cite{Puri-Blais-2017}. Indeed, engineering a nonlinear Hamiltonian of the form
\begin{align*}
H_{\text{kerr}}&=-K\ba^{\dag 2}\ba^2+\epsilon_{2\text{ph}}\hat{a}^{\dag 2} + \epsilon^*_{2\text{ph}}\hat{a}^2\\
&=-K(\ba^{\dag 2}-\alpha^{*2})(\ba^{ 2}-\alpha^{2}),
\end{align*}
with $\alpha=\sqrt{\epsilon_{2\text{ph}}/K}$ the ground states $\ket{\pm\alpha}$,  are twofold degenerate. The system can be thought of as a double well potentiel, where the tunneling between the two wells is exponentially suppressed with $|\alpha|^2$. Note, however, that, with such a Hamiltonian protection, some type of friction needs to be added in order to avoid leakage errors (out of the encoded qubit subspace) due to excursions in each well. The natural photon loss of the harmonic oscillator, if it is stronger than the mechanisms leading to such an excursion, can compensate this leakage. Thus, a  promising  approach is a combination of the Kerr type Hamiltonian and two-photon dissipation, where the protection against leakage does not come at the expense of higher phase-flip rates.

In the next section, we show how to extend this half-protection to a full protection against both phase flips and bit flips. More precisely, we design an economic encoding that suppresses the phase flips without reintroducing bit flips.

\section{From cat qubits to  protected logical qubits}\label{sec:overall_scheme}

\bgroup
\def\arraystretch{2.5}
 \begin{table*}[t]
 \begin{center}
    \begin{tabular}{| >{\centering\arraybackslash}p{5,8cm}  | >{\centering\arraybackslash}p{6.5cm} | >{\centering\arraybackslash}p{4.5cm}  |}
    \hline
    Physical qubits & Two-level systems with biased noise~\cite{Aliferis2008, Webster2015} & Cat-qubits \\ \hline
    Fundamental bias-preserving operations & $\cG_0=\{ \mathcal{P}_{\ket+}, \mathcal{M}_{X}, \text{CPHASE}, Z(\theta), \text{CZ}(\theta)  \}$ &  $\cG_0 \cup\{X, \text{CNOT}, \text{Toffoli} \} $\\ \hline
    $\Cone$-logical operations &  \parbox[c]{6cm}{$\cG_1=\{ \mathcal{P}_{\ket0}, \mathcal{P}_{\ket+}, \mathcal{M}_{X}, \mathcal{M}_{Z}, \text{CNOT}  \}$ } &  $\cG_1\cup\{X, \text{Toffoli}\}$, 
 \textbf{Universal}   \\[8pt] \hline
    $\Cone \triangleright \Ctwo$-logical operations & \parbox[c]{6cm}{$\cG_2=\cG_1\cup \{\mathcal{P}_{|+i\rangle}, \mathcal{P}_{|T\rangle}\}$, \textbf{Universal} 
}  & - \\[5pt] \hline
    \end{tabular}
  \caption{\label{RegularVsCats} Construction of a universal set of fault-tolerant gates that exploit the noise bias of the physical qubits. {The middle column represents the case of regular two-level systems~\cite{Aliferis2008, Webster2015}, and the right column represents the case of cat qubits. } For regular qubits, only a few fundamental bias-preserving operations are allowed, leading to a limited set of $\Cone$-logical operations. To achieve universality, it is necessary to concatenate with a second level of encoding $\Ctwo$, at which magic state preparation and distillation are appended. On the other hand, the set of fundamental biased-preserving operations for cat qubits contains extra gates. {This extended set enables us to build a universal set of fault-tolerant gates already at the repetition code level, and this without requiring magic state preparations and distillations. Furthermore, the circuits for realization of protected logical gates are significantly simpler than regular two-level systems. }}
  \end{center}
 \end{table*}
\egroup

We see that cat qubits admit a biased noise where the bit-flip errors are suppressed exponentially with the cat size. In this section, we trace out a viable path toward full protection with minimal hardware overhead.

It is tempting to think that physical qubits suffering only from phase-flip errors can be protected  through a simple classical-error correction-scheme such as a repetition code. All that is required is the ability to perform parity-type measurements between neighboring qubits. Indeed, this idea can be explored to build a fully protected quantum memory, but performing protected logical gates comes with further complications. The main issue  is that, the execution of a gate can in principle convert a phase-flip error into a bit-flip one, which is not suppressed by the simple error correction. One is therefore limited to only employ physical operations that preserve the noise bias (i.e. do not convert phase flips into bit flips). Such operations are called \textit{bias preserving.} 

This idea is employed as a first level of encoding in Ref.~\cite{Aliferis2008, Webster2015}. In these papers, the quantum information is protected by a concatenation of two codes $\mathcal{C}_1 \triangleright \mathcal{C}_2$, where $\triangleright$ denotes code concatenation. The code $\mathcal{C}_1$ is a length-$n$ repetition code that protects against phase flips errors, producing logical qubits that suffer from an effective unbiased noise of strength $\epsilon_1$. As soon as $\epsilon_1$ is below the threshold of $\mathcal{C}_2$, arbitrarily low logical error rates $\epsilon_2$ can be achieved by the concatenated code. This second level of encoding is required even if the qubits do not suffer from bit-flip errors at all, that is in the limit of an infinite noise bias. {Indeed,} even if the logical error rate $\epsilon_1$ of the repetition code $\Cone$ can be made arbitrarily low, it is not possible to build a universal gate set for the $\Cone$-encoded logical qubits, using only bias preserving  operations. In this paper, we see that this no-go theorem is broken by using cat qubits (instead of regular two-level systems) as base qubits of the repetition code.  

As discussed in Refs.~\cite{Aliferis2008,Webster2015}, some operations are naturally bias preserving.  In the case of dominant phase-flip errors, the preparation of $\ket{\pm}=(\ket{0}\pm\ket{1})/\sqrt 2$ states $\cP_{\ket{\pm}}$, and  the measurement of the $X$ operator $\cM_X$, are bias preserving because the eigenstates $\ket{\pm}$ of the $X$ operator ($\sigma_x$ Pauli operator) are fixed points for bit-flip errors. Also, the controlled phase gates $\text{CPHASE} = \tfrac12 (I_1 + Z_1)\otimes I_2 + \tfrac12(I_1-Z_1)\otimes Z_2$ (with $Z_j$ being the Pauli $\sigma_z$ operator on qubit $j$), or more generally the two-qubit entangling gate CZ$(\theta)=\exp(i\theta/2 Z_1\otimes Z_2)$~\cite{Webster2015}, can be  implemented in a bias-preserving manner. Indeed, it is enough to note that the Hamiltonians proportional to $Z_1\otimes Z_2-Z_1/2-Z_2/2$ or  $Z_1\otimes Z_2$ that generate such unitary operations commute with phase-flip errors.  
However, a  universal gate set  will necessarily contain gates that do not commute with the dephasing errors, such as  CNOT or  Toffoli. 

{Note that, while these gates do not commute with the phase-flip errors, their action may still be compatible with noise bias. Considering the CNOT$=\tfrac12 (I_1 + Z_1)\otimes I_2 + \tfrac12(I_1-Z_1)\otimes X_2$ gate for instance, $Z_1$ errors commute with CNOT and $Z_2$ errors are converted into correlated phase-flip errors $Z_1Z_2$.} {This overall action does not  to convert phase flips to bit flips and is therefore correctable by the repetition code.} However, as {correctly} noted by~\cite{Aliferis2008}, the same property is not necessarily satisfied \textit{during the execution} of the gate. {Indeed, we  prove in the Appendix~\ref{append:nogo} that such a conversion of phase flips to bit flips necessarily occurs when implementing a CNOT gate with two-level systems. In other words, the CNOT operation cannot be performed in a bias-preserving manner with two-level systems. The same result also holds for a Toffoli gate. }

\begin{figure*}[t]
\includegraphics[width=.9\linewidth]{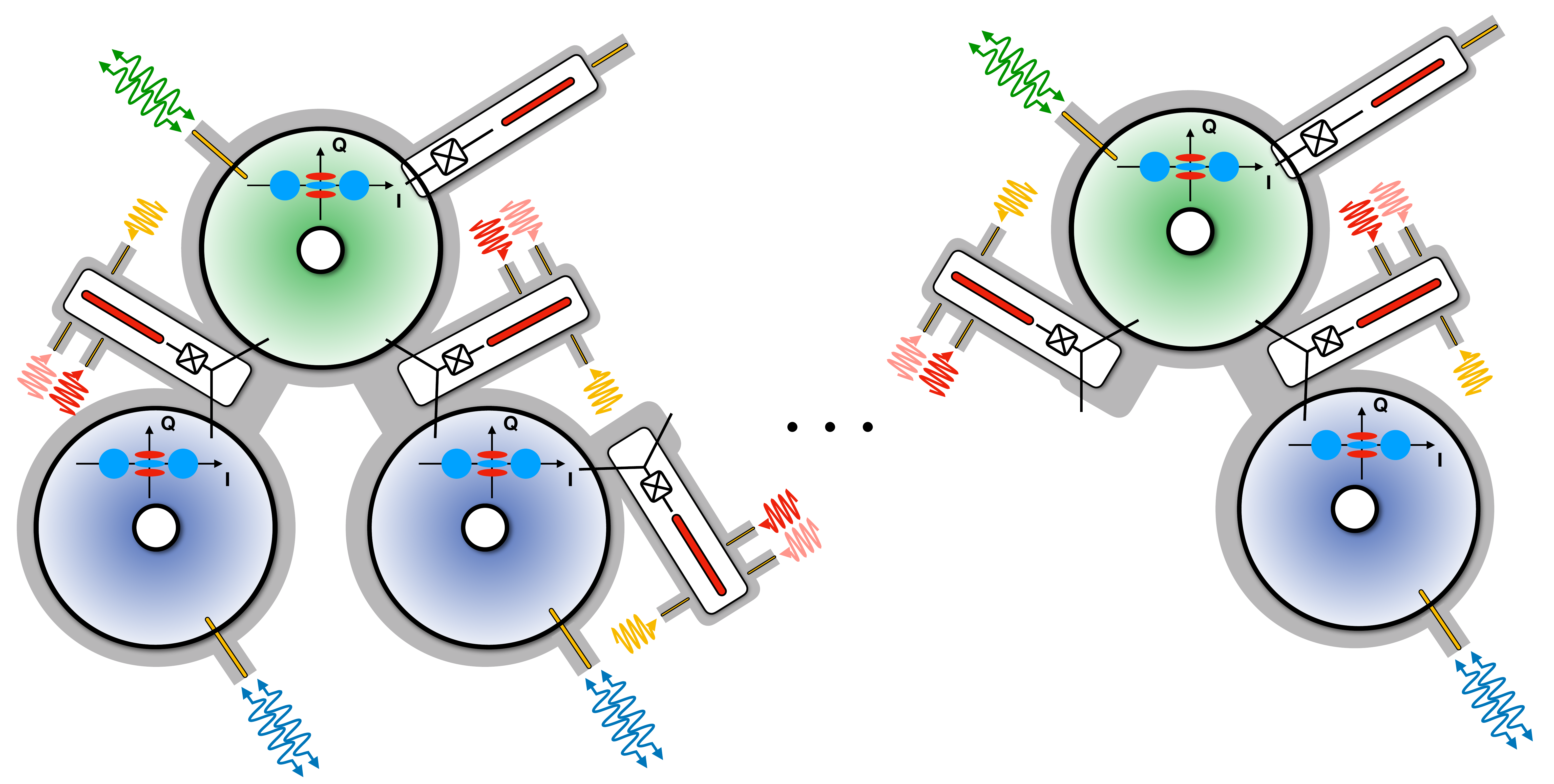}
\caption{\label{RepetitionCat} {Layout of a repetition cat qubit using high-Q 3D cylindrical postcavities~\cite{Reagor-PRB-2016}. Each data cat qubit (in blue cavities) is connected to a pair of ancilla cat qubits (in green cavities) for the joint parity measurement. The results of the parity measurement are read out using the low-Q stripline resonators (in red) coupled to green cavities. Each cat qubit is continuously driven via the two-photon driven-dissipative scheme (arrows). The couplings between  cavity modes are mediated by a Josephson circuit and extra microwave drives, required for bias preserving CNOT operations as detailed in Sec.~\ref{sec:physical_gates} and Fig.~\ref{fig:CNOT_Toffoli_exp}. The choice of cylindrical postcavities is to ensure high quality factors, but a similar layout could be thought of in a 2D architecture.}}
 \end{figure*}

In Ref.~\cite{Aliferis2008}, the set of bias-preserving operations is therefore limited to $\{\mathcal{P}_{\ket+}, \mathcal{M}_{X}, \text{CPHASE}  \}$, which is not enough to build a universal gate set at the level of the repetition code $\Cone$. The concatenation with $\Ctwo$ is therefore necessary to gain universality. At the $\Cone \triangleright \Ctwo$-logical level, the full set of Clifford gates can be achieved by preparing the magic state $\ket{+i}_L := \tfrac{1}{\sqrt2}(\ket0_L + i\ket1_L)$, and this set becomes universal with the addition of another magic state $\ket T_L := \tfrac{1}{\sqrt2}(\ket0_L + e^{i\pi/4}\ket1_L)$. In Ref.~\cite{Webster2015}, with the addition of CZ$(\theta)$ to the set of bias-preserving operations, the authors construct new gadgets to reduce the overhead for magic state preparation and distillation. The construction of Refs.~\cite{Aliferis2008,Webster2015} to exploit a noise bias in regular qubits (two-level systems) is summarized in Table~\ref{RegularVsCats}, in order to clarify the radical simplification due to the use of cat qubits.

In this paper, we show that the cat qubits have specific features which allow us to circumvent {the aforementioned obstacles and significantly reduce the complexity of protected logical gates. These features rely on  the infinite-dimensional Hilbert space of the oscillator in  which the two-dimensional Hilbert space of the cat qubit is embedded}.  More precisely, gates are performed by a continuous distortion of the two-dimensional manifold defining the cat qubit, in such a way that the {exponential suppression of bit flips remains valid} during the execution of the gate. The apparent ``magic'' comes from the fact that the $Z$ component of the qubit is transformed continuously, which would not be possible using a DV system. Following this idea, we detail in Sec.~\ref{sec:physical_gates} how a universal set of bias preserving gates can be implemented at the cat qubit level. Even more remarkably, the realization of this set requires hardware-efficient operations, with {no} use of functional ancilla qubits, nor magic state preparation, distillation and injection. 

In order to extend the protection to phase-flip errors, we propose to embed the cat qubits in a repetition code (Fig.~\ref{RepetitionCat}). This repetition code $\mathcal{C}_1$ is defined in the dual basis. The code space is defined  as the +1 common eigenspace of the $n-1$ stabilizers
$$
S_j =X_j   X_{j+1}, \qquad j \in [\![1,n-1]\!].
$$
The logical operators for the \textit{repetition cat qubit} are 
 $$
 X_L = X_1, \hspace{8pt} Z_L = \bigotimes \limits_j Z_j, \hspace{8pt}  Y_L = i X_L Z_L.  
 $$
 The logical $\ket+_L$ and $\ket-_L$ states are given by $\ket\pm_L := \ket\pm_c^{\otimes  n} = \ket{\mathcal{C}_\alpha^\pm}^{\otimes n}$. Note that this definition leads to the following nontrivial logical computational states: 
 \begin{align*}
 \ket0_L &=  \tfrac{1}{(\sqrt2)^{n-1}} \sum \limits_{j \in \{0,1\}^n, |j|\ \text{even}}\ket j_c \\
 \ket1_L &=  \tfrac{1}{(\sqrt2)^{n-1}} \sum \limits_{j \in \{0,1\}^n, |j|\ \text{odd}}\ket j_c
 \end{align*}
 where $j$ is an $n$-bit string composed of 0's and 1's and $|j|$ denotes the number of 1's, called the \textit{weight}, of the string $j$. Recalling that $\ket1_c \approx \ket{-\alpha}$, one can note that the logical information is encoded in the parity of the number of oscillators in the $\ket{-\alpha}$-state.
 
 This code, {with phase-flip error correcting capacity of $(n-1)/2$}, does not detect nor correct physical bit-flip errors. Here, $n$ is chosen such that the probabilities $p_{Z_L}$ of logical phase-flip,  and $p_{X_L}$ of  logical bit-flip errors, are of comparable strength, thus producing a $\Cone$-logical qubit suffering from an effective \textit{unbiased} noise of strength ${\epsilon_L} := p_{X_L} + p_{Z_L}$. It is worth noting that  the accuracy that can be achieved by $\mathcal{C}_1$ is set by the size of the cat states. Indeed, the lower bound for this accuracy is given by $p_{X_L}$ and decreases exponentially to zero as the cat size increases. Fixing a reasonable mean  photon number of $\bar n=10$, this achievable accuracy could be as low as $10^{-9}$ and reduces to $10^{-13}$ for $\bar n=15$. Cat states of such sizes have been previously prepared in the context of superconducting circuits~\cite{vlastakis-science-2013}.  If for unforeseeable reasons, the locality assumption on the noise processes breaks down for larger cats, one can consider a concatenation with a second code $\mathcal{C}_2$ to achieve even better accuracies. Note, however, that this second level of concatenation could be done with any simple low-order code to go from already small error probabilities (e.g., $10^{-9}$) down to the required precision for a given algorithm.

 \begin{figure*}[t!]
\includegraphics[width=\linewidth]{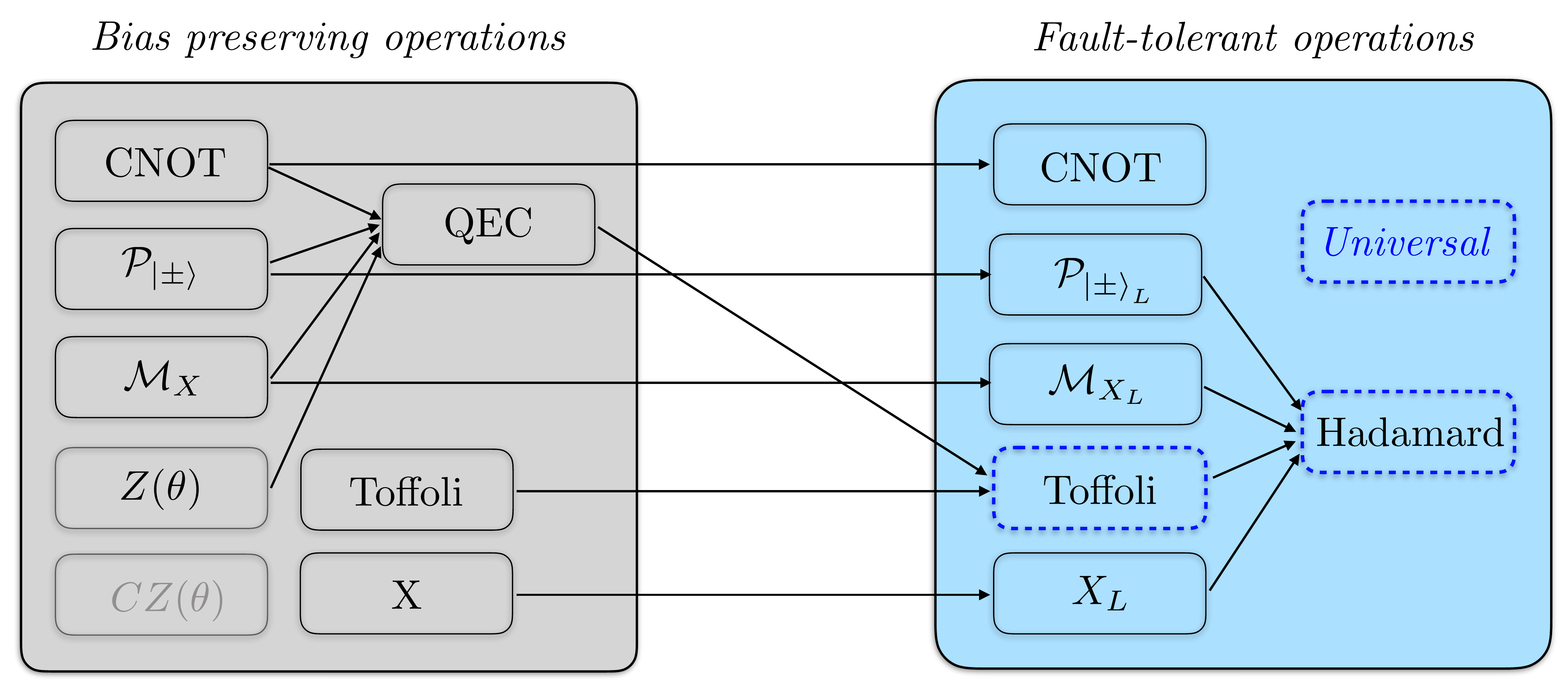}
\caption{\label{Scheme} Overall scheme for achieving fault-tolerant universal quantum computation using cat qubits. The fundamental operations (left-hand side) are performed on the cat qubits, in a bias-preserving manner. Fault-tolerant logical operations acting on the repetition cat qubits (right-hand side) are built out of these operations, as depicted by the arrows. This construction is detailed in Sec.~\ref{sec:logical_gates}}.
 \end{figure*}
 
Let us now briefly present the picture for the construction of fault-tolerant gates at the level of the repetition cat qubits (see~Fig.~\ref{Scheme}). As discussed earlier, the  set of bias-preserving fundamental operations at the cat qubit level includes $\mathcal{S} = \{\mathcal{P}_{\ket\pm_c}, \mathcal{M}_X, X, Z, \text{CNOT}, \text{Toffoli}\}$ (see Sec.~\ref{sec:physical_gates} for a detailed construction). The next step is to build fault-tolerant encoded operations at the level of the repetition code, using operations from this fundamental set. The set of fault-tolerant encoded operations acting on repetition cat qubits is given by $\mathcal{S}_L = \{\mathcal{P}_{\ket\pm_L},  \mathcal{M}_{X_L}, X_L, \text{CNOT}, \text{Toffoli}\}$. It is worth noting that the logical CNOT can be implemented transversally from the fundamental CNOT, (see Sec.~\ref{sec:logical_gates} for the details), leading to great hardware simplifications when compared to the logical CNOT construction of~\cite{Aliferis2008}.
The universality of $\mathcal{S}_L$ is established by the fact that it contains the Toffoli gate in the computational basis, while state preparation and measurement are done in the dual basis. Indeed, we show in~Sec.~\ref{sec:logical_gates} how a logical Hadamard gate (single-qubit basis-changing operation) can be built out of the gates of the set $\mathcal{S}_L$, thus achieving universality~\cite{Shi2003, Aharonov-2003}.

\section{Bias-preserving operations}\label{sec:physical_gates}
 In this section, we first explain how the set of operations $\mathcal{S} = \{\mathcal{P}_{\ket\pm_c}, \mathcal{M}_X, X, Z, \text{CNOT}, \text{Toffoli}\}$ can be realized at the cat qubit level in a bias-preserving manner. The operations in $\mathcal{S}$ are sufficient to build the universal set of logical gates for the repetition code (see Sec.~\ref{sec:logical_gates}). In addition, we recall at the end of the section how arbitrary rotations around the $Z$ axis $Z(\theta)$ (proposal~\cite{Mirrahimi2014}, experimental realization~\cite{Touzard-PRX-2018}) and   the two-qubit entangling gate  CZ$(\theta)=\exp(i\theta/2 Z_1\otimes Z_2)$~\cite{Mirrahimi2014} can also be realized. Even if these operations are not needed for the theoretical construction of this paper, they may prove useful for an optimized implementation of quantum algorithms.

\paragraph*{Preparation of $\ket\pm_c$.}
First, we note that the states $\ket{\pm}_c$ are eigenstates of the logical $X$ operator which make their preparation compatible with the noise bias (suppressed bit flips)~\cite{Aliferis2008}. The preparation of the even cat state  $\ket+_c = \Cp$ is performed merely by turning on the driven two-photon dissipation~\eqref{eq:2photon}, when the system is initialized in the vacuum state $\rho(0) = \ket0\bra0$~\cite{Mirrahimi2014}.  Indeed, the conservation of photon-number parity due to the two-photon driven dissipation ensures that the steady state of the system is given by the even cat state. Such a state preparation has already been realized experimentally ~\cite{Leghtas2015} {and the fidelity of the operation is set by the ratio between the two-photon dissipation rate $\kappa_{2\text{ph}}$, setting the rate of convergence to the cat state, and the undesired single-photon loss rate $\kappa_{1\text{ph}}$, setting the parity jump rates (equivalent to phase-flip errors) mixing the even cat with the odd one. In the latest experiments a ratio of about $\kappa_{1\text{ph}}/\kappa_{2\text{ph}}=10^{-2}$ has been achieved between these two rates~\cite{Touzard-PRX-2018} and further improvements seem to be within the reach of the current experiments.

A systematic way to prepare the odd cat state $\ket{-}_c=\Cm$ is to start with preparing the even cat state and then performing a $Z$ operation. A bias preserving rotation around the $Z$ axis is proposed in Ref.~\cite{Mirrahimi2014} and experimentally realized in Ref.~\cite{Touzard-PRX-2018}. For the sake of completeness, we recall the idea behind this realization at the end of this section. While this proves the feasibility of the physical preparation of $\ket{-}_c$, at the end of Sec.~\ref{sec:logical_gates}, we show that, in practice, one can replace such a $Z$ operation by a simple $Z$ operation in classical software~\cite{Knill2005}. This process reduces the number of physically implemented logical gates in an algorithm.

Finally, we also note that such a state preparation can be performed through other strategies as well. In particular, in many recent experiments (e.g., Ref.~\cite{Touzard-PRX-2018}), these states are generated using optimal control techniques which can significantly improve the  fidelity with respect to a passive preparation with two-photon driven dissipation. 

\paragraph*{Measurement of $X$.}
For the purpose of our scheme, the measurement of $X$ (photon-number parity measurement) could be either destructive or quantum nondemolition (QND) as it is always used on ancilla qubits which can be discarded after each measurement. However,  a QND protocol allows us to achieve a better fidelity by repeating the measurements. The QND parity measurement proposed in Ref.~\cite{Lutterbach-Davidovich-97} and realized in Refs.~\cite{Bertet-PRL-2002,Sun2014} is a perfectly valid protocol for our scheme. For the sake of completeness, we recall the main idea behind this measurement protocol. The cavity whose parity is to be measured is coupled to an ancilla qubit via the dispersive interaction Hamiltonian 
\[
\hat{H}_{\text{disp}} = -\chi \ket e \bra e \hat{a}^\dag\hat{a}.
\]
The evolution on a time interval $T=\pi/\chi$ is given by the unitary 
\[
\hat{\mathcal{U}} = \ket g \bra g   I + \ket e \bra e   e^{i\pi\hat{a}^\dag\hat{a}}
\]
entangling the state of the ancilla with the parity of the state of the cavity. Preparing the ancilla qubit in a superposition state $\ket + = \tfrac{1}{\sqrt{2}}(\ket g + \ket e)$, the effect of the unitary $\hat{\mathcal{U}}$ is to flip the ancilla to the state $\ket - = \tfrac{1}{\sqrt{2}}(\ket g - \ket e)$ when the cavity contains an odd number of photons and to leave it unchanged otherwise. A measurement of the $\hat{\sigma}_x$ operator of the qubit thus reveals the parity of the cavity state. 

{Note that, in order to perform such a parity measurement, we need to turn off the two-photon driven dissipation on the measured system. However, as stated earlier, these measurements are performed on ancilla cavities that are thrown out after each measurement. So the absence of protection during the measurement affects merely the measurement fidelity and does not have any consequence on the rest of the circuit. Rather high parity-measurement fidelities of about 98.5\% have been previously achieved using this protocol~\cite{Ofek-Petrenko-Nature-2016}.}

\paragraph*{$X$ gate.}
Our realization of the $X$ gate is  based on an adiabatic deformation of the code space. As discussed in Sec.~\ref{sec:bias}, the effective dissipation channel $\kappa_2\mathcal{D}[\hat{a}^2 - \alpha^2]$ stabilizes the two-dimensional subspace $\text{span}\{\ket{\pm\alpha} \}$. It is possible to perform nontrivial operations on the encoded information by varying the complex number $\alpha$ in time. When the variations of $\alpha(t)$ are sufficiently slow with respect to $\kappa_2^{-1}$, the dissipator $\kappa_2\mathcal{D}[\hat{a}^2 - \alpha(t)^2]$ stabilizes $\text{span}\{\ket{\alpha(t)}, \ket{-\alpha(t)} \}$ at all times t. This stabilization should be thought of as a slow motion of the fixed points of the dynamics in the phase space.

Remarkably, such a deformation preserves the quantum information, provided the two states $\ket{\alpha(t)}$ and $\ket{-\alpha(t)}$ remain sufficiently separated in phase space at all times: The state $\ket{\psi_0} = c_0 \ket\alpha + c_1\ket{-\alpha}$ at time $t=0$ evolves under the effect of $\kappa_2\mathcal{D}[\hat{a}^2 - \alpha(t)^2]$, with $\alpha(0)=\alpha$, to $\ket{\psi_t} = c_0\ket{\alpha(t)}+ c_1\ket{-\alpha(t)}$ provided $|\dot{\alpha}(t')|/|\alpha(t')| \ll \kappa_2$ and $|\braket{\alpha(t')}{-\alpha(t')}|^2 \ll 1$ at all times $t' \in [0,t]$.  

An $X$ operation can be realized in such a manner by choosing a "path" function $\alpha(t)$ such that $\ket{\alpha}$ and $\ket{-\alpha}$ are exchanged, e.g $\alpha(t) = \alpha e^{i \pi t/T}$, $t \in [0, T]$ where $T \gg \kappa_2^{-1}$ is the gate time~\cite{Puri2019}. Indeed, the swap $\ket\alpha \leftrightarrow \ket{-\alpha}$ corresponds to the map $\Cp \rightarrow \Cp$ and $\Cm \rightarrow -\Cm$ which is an $X$ operation for the cat qubit. In addition to such a topological phase, there is a geometric phase accumulated due to the particular path taken by $\alpha(t)$. However, this phase is the same for the two states $\ket{\pm\alpha}$ and corresponds to a global phase.

In the ideal case of a lossless harmonic oscillator  and in the limit where the gate time $T = +\infty$, the fidelity of this operation with respect to the X operator is 1. This operation is bias preserving, as the errors caused by the finite gate time are only of the phase-flip type, but the bit flips remain exponentially suppressed in the size of the cat $\bar n$. Intuitively, this result is not surprising, as the two-photon pumping is never turned off during the gate execution.  We depict in Fig.~\ref{fig:rotating_cat} a schematic representation of this evolution in the phase space.

\begin{figure}[h]
\includegraphics[width=.5\textwidth]{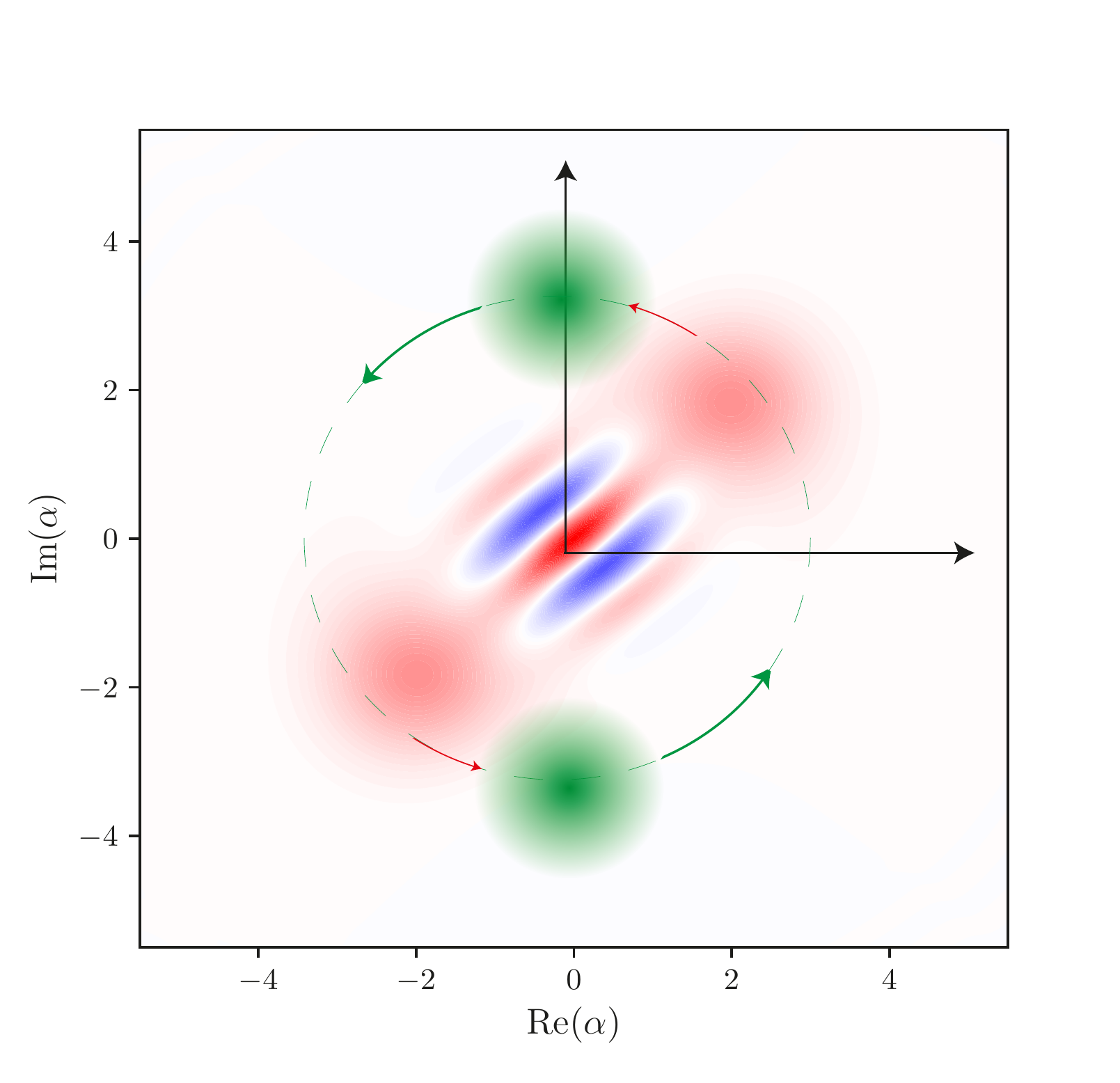}
\caption{\label{fig:rotating_cat} Wigner function of the state of a cat qubit during the execution of an $X$ operation. The green dots are the Wigner functions of the instantaneous steady states of the dynamics $\dot\rho =  \kappa_2\mathcal{D}[\hat{a}^2 - \alpha(t)^2]$. These attractive points are slowly rotated from $\pm \alpha$ to $\mp \alpha$ on the dashed circle, as shown by the green arrows. When this rotation is performed slowly, the cat follows the attractors (red arrows).
}
\end{figure}

To reduce the phase-flip error rate due to the finite gate time (nonadiabaticity) , the Hamiltonian $\hat H = - \frac\pi T \hat{a}^\dag\hat a$ is turned on while the pumping is being rotated. This Hamiltonian generates the unitary $\hat{\mathcal{R}}(t) = e^{i\frac \pi T \hat{a}^\dag\hat a t}$ which rotates deterministically the qubit state $\hat{\mathcal{R}}(t) \ket{\psi_0}  = c_0\ket{\alpha(t)}+ c_1\ket{-\alpha(t)}$ so that it remains at all times a pointer state of the time-dependent dissipative channel:
\[
[\hat{a}^2 - \alpha(t)^2] \hat{\mathcal{R}}(t) \ket{\psi_0} = 0
\]
In the presence of this Hamiltonian, there is no need to proceed adiabatically; that is, the gate time $T$ can be arbitrarily short.

\paragraph*{CNOT gate.} 
The idea described above can be  adapted to realize a $\text{CNOT} = \tfrac12 (I_1 + Z_1)  I_2 +  \tfrac12 (I_1 - Z_1)  X_2$.
The CNOT for the cat qubits is given by
\[
\text{CNOT} \approx  \ket\alpha\bra\alpha \otimes I_\alpha + \ket{-\alpha}\bra{-\alpha} \otimes X_\alpha
\]
where $I_\alpha =  \ket\alpha\bra\alpha + \ket{-\alpha}\bra{-\alpha}$ and $X_\alpha = \ket{\alpha}\bra{-\alpha} + \ket{-\alpha}\bra{\alpha}$. {The approximation is exponentially precise in $|\alpha|^2$.} Inspired by the proposal of Ref.~\cite{Puri2019} for the Kerr cats, this operation is realized by making the rotation of the pumping of the target qubit (implementing $X_\alpha$; see the previous paragraph) conditional to the state of the control qubit. In our case, this operation is realized in time T by the two dissipation channels $\mathcal{L}_{\hat{a}} = \mathcal{D}[\hat L_{\hat a}]$ and $\mathcal{L}_{\hat{b}} = \mathcal{D}[\hat L_{\hat b}(t)]$, with:
\begin{equation*}
\hat L_{\hat a} = \hat{a}^2 - \alpha^2,\qquad
\hat L_{\hat b}(t) = \hat{b}^2 - \tfrac12\alpha(\hat{a}+\alpha) + \tfrac12\alpha {e^{2 i \frac \pi T t}}(\hat{a}-\alpha)
\end{equation*}
 where we denote by $\hat{a}$ (respectively, $\hat{b}$) the mode of the control cat qubit (respectively, target cat qubit). The dissipation channel on the control qubit $\mathcal{L}_{\hat{a}}$ is the two-photon pumping scheme stabilizing the control cat qubit. The second dissipation channel, however, acts on the target cat qubit but also depends on the first mode $\hat a$. It should be understood as follows: When the control qubit $\hat a $ is in the state $\ket\alpha$, the operator $\hat L_{\hat b}(t)$ acts on the target mode as $\hat{b}^2 - \alpha^2$, stabilizing the idle code space, but when the control qubit is in the state $\ket{-\alpha}$, the pumping becomes $\hat{b}^2 - (\alpha e^{ i \frac \pi T t})^2$,  thus implementing the time-dependent two-photon pumping dissipation used for the $X_\alpha$ operation. Again, the pumping is never turned off and the bit-flip errors remain exponentially suppressed at all times, thus ensuring that the CNOT gate preserves the biased structure of the noise.  In Sec.~\ref{sec:experimental}, we explain how the experimental realization of such a time-dependent dissipation operator is a straightforward modification of the regular two-photon driven dissipation~\cite{Leghtas2015}.
 
We now explain how to deal with two undesired effects that limit the fidelity of the operation: the geometric phase due to the path taken by the states $\ket{\pm\alpha(t)}$ in the phase space, and the phase-flip errors induced by the finite gate time (nonadiabaticity). 
In the case of the $X$ gate, the geometric phase corresponded to a physically meaningless global phase, but here this phase is conditioned on the state of the control qubit. As a consequence, the geometric phase induces a deterministic rotation around the $Z$ axis of the control qubit. The rotation angle is given by 
$$
\vartheta=-i\int_0^T \bra{\pm\alpha(t)}  \frac{d}{dt}\ket{\pm\alpha(t)}dt=\pi|\alpha|^2.
$$
This deterministic geometric phase can be compensated by {applying a local $Z(\theta)$ operation (see below). A second option is to ensure that the rotation angle $\vartheta$ is a multiple of $2\pi$, either by setting the number of photons to be an even integer or by choosing a path $\alpha(t)$ such that the result of the integral is a multiple of $2\pi$. Even in this case, the fluctuations along the chosen path inevitably lead to a certain imprecision in the final value of the geometric phase. This situation is not an issue, as it can lead only to phase-flip errors, accounted for by the repetition code.}

A major part of the phase-flip errors induced by nonadiabatic effects can be compensated in the same way as the X operation, by adding a Hamiltonian evolution of the form
\[
\hat H = \frac12 \frac \pi T \frac{ \hat a -\alpha}{2\alpha} \otimes (\hat{b}^\dag \hat b-\bar n) + H.c.
\]
while rotating the pumping. In the presence of two-photon pumping, this Hamiltonian  is an approximation of $\pi/T \ket{-\alpha}\bra{-\alpha}\otimes (\bb^\dag\bb-\bar n)$, rotating the target cat qubit conditional to the control cat qubit being in the state $\ket{-\alpha}$. Such Hamiltonians have been already realized using parametric methods~\cite{Touzard-PRL-2019}, similar to those used in driven two-photon dissipation. 

\paragraph*{Toffoli gate.}
The Toffoli gate is the three-qubit gate corresponding to a "controlled-controlled-NOT": 
\begin{multline*}
\text{Toffoli} = \tfrac14 (I_1 + Z_1)  (I_2 + Z_2)   I_3 +  \tfrac14 (I_1 + Z_1)  (I_2 - Z_2)   I_3 \\
+  \tfrac14 (I_1 - Z_1)  (I_2 + Z_2)   I_3 + \tfrac14 (I_1 - Z_1)  (I_2 - Z_2)   X_3.
\end{multline*}
This unitary does not belong to the Clifford group. In fact, this gate, together with any set of gates generating the Clifford group, is  universal. In the vast majority of schemes achieving universality, the non-Clifford operation is by far the most difficult operation to be realized. A remarkable feature of our scheme is that the physical implementation of the Toffoli at the cat qubit level is very much like the CNOT gate and, thus, of similar complexity. Three dissipation channels are realized, $\mathcal{L}_{\hat{a}} = \mathcal{D}[\hat L_{\hat a}]$, $\mathcal{L}_{\hat{b}} = \mathcal{D}[\hat L_{\hat b}]$ and $\mathcal{L}_{\hat{c}} = \mathcal{D}[\hat L_{\hat c}(t)]$,
\begin{align*}
\hat L_{\hat a} &= \hat{a}^2 - \alpha^2,\qquad
\hat L_{\hat b} = \hat{b}^2 -\alpha^2\\
\hat L_{\hat c}(t) &= \mathcal{D}[\hat{c}^2 -\tfrac14 (\hat a + \alpha)(\hat b + \alpha)
+\tfrac14 (\hat a + \alpha)(\hat b - \alpha) \\
& +\tfrac14 (\hat a - \alpha)(\hat b + \alpha)- \tfrac14 e^{{2}i \frac \pi T t} (\hat a - \alpha)(\hat b - \alpha) ].
\end{align*}
Here, {$\mathcal{L}_{\hat{a}}$ and  $\mathcal{L}_{\hat{b}}$ keep stabilizing the two control modes $\hat a$ and $\hat b$ in manifolds spanned by $\ket{\pm\alpha}$, and  $\mathcal{L}_{\hat{c}}$ rotates the two-photon pumping on the target mode $\hat c$ only when the control cat qubits are in the state $\ket{-\alpha, -\alpha}$. As for the CNOT gate, two effects (the geometric phase and the nonadiabaticity) limit the gate fidelity. The deterministic geometric phase associated to the path taken by the target cat qubit can also be eliminated by tailoring the path followed in the phase space by the cat states during the execution of the gate, or by physically applying $Z(\theta)$ and $CZ(\theta)$ operations. To reduce the phase-flip errors induced by non adiabaticity, the Hamiltonian 
 \[
 \hat H = -\frac12 \frac \pi T \frac{\hat a-\alpha}{2\alpha}\otimes \frac{\hat b-\alpha}{2\alpha}\otimes {(\hat{c}^\dag \hat c - \bar n)} + h.c.
 \]
 is added. We analytically analyze the performance of this gate in Sec.~\ref{sec:error_analysis}, and in Sec.~\ref{sec:experimental}, we discuss an experimental implementation. 

From a theoretical point of view, assuming the required couplings between any number of modes are available, the mechanism presented above could be straightforwardly extended to realize the $n$-qubit entangling gate $C^{n-1} X$ where $C^{n-1}$ denotes the controls on the first $n-1$ qubits (the CNOT being $CX$ and the Toffoli, $C^2 X$). 

\paragraph*{Rotation around Z of an angle $\theta$.}
As discussed in Ref.~\cite{Mirrahimi2014} and  experimentally realized in Ref.~\cite{Touzard-PRX-2018}, the quantum Zeno effect can be used to perform a rotation of an arbitrary angle $\theta$ around the $Z$ axis in a bias-preserving manner:
\[
Z(\theta) = \cos\frac{\theta}{2} I_\alpha + i \sin\frac{\theta}{2} Z_\alpha
\]
where $I_\alpha = \Cp \Cpd + \Cm \Cmd $ and $Z_\alpha =  \Cp \Cmd + \Cm \Cpd$. To do so, a weak resonant drive $\hat H = \epsilon_Z (\ba + \ba^\dagger)$ is applied in the presence of the two-photon driven dissipation. When the single-photon drive is much weaker than the two-photon dissipation, $\epsilon_Z \ll \kappa_{2\text{ph}}$, it induces effective oscillations in the equatorial plane of the Bloch sphere whose frequency is given by $\Omega_Z = 2\epsilon_Z |\alpha|$. 

\paragraph*{Two-qubit entangling gate $CZ(\theta)$.}
In the same spirit, a bias preserving two-qubit entangling gate 
\[
CZ(\theta) = \cos\frac{\theta}{2} I_1 I_2 + i\sin\frac{\theta}{2} Z_1 Z_2
\] 
can be implemented using a weak beam-splitter Hamiltonian $\hat H = \epsilon_{ZZ} (\ba_1\ba_2^\dagger + \ba_1^\dagger \ba_2)$ in the presence of the two-photon driven dissipation (see Ref.~\cite{Mirrahimi2014} for more details).

 \section{Universal set of logical gates}\label{sec:logical_gates}
 In this section, we construct the fault-tolerant logical operations of the set $\mathcal{S}_L =  \{\mathcal{P}_{\ket\pm_L}, \mathcal{M}_{X_L}, X_L, \text{CNOT}_L, \text{Toffoli}_L \}$ for the repetition code, using the fundamental set of bias-preserving operations $\mathcal{S} = \{\mathcal{P}_{\ket\pm_c}, \mathcal{M}_{X}, X, Z, \text{CNOT}, \text{Toffoli}\}$ for individual cat qubits. The universality of $\mathcal{S}_L$ is established by building a Hadamard gate out of operations in this set, which, together with the Toffoli gate, is universal~\cite{Shi2003,Aharonov-2003}.  We start by explaining how the quantum error correction of the repetition code is realized.

 \paragraph*{Quantum error correction for  repetition code.}
A prerequisite for using a repetition code is the ability to measure the value of the $n-1$ stabilizers of the code, also called the \textit{parity-check operators}. In our setup, these stabilizers are the joint-parity operators of any two pairs of neighboring cat qubits $X_jX_{j+1}$. The measurement of these operators can be achieved using operations in $\mathcal{S}$, with the circuit in Fig.~\ref{fig:XX_measurement}.
\begin{figure}[h!]
\[
\Qcircuit @C=.7em @R=.4em @! {
\lstick{\ket{\psi_{j}}_c}   & \targ & \qw  & \qw \\
\lstick{\ket{\psi_{j+1}}_c}  & \qw & \targ  & \qw \\
\lstick{\ket{+}_c} & \ctrl{-2} & \ctrl{-1}  & \gate{\mathcal{M}_X} & 
}
\]
\caption{
Joint-parity measurement between two neighboring  cat qubits $j$ and $j+1$ of a repetition cat qubit, using one ancilla cat qubit. Note that the error propagation from ancilla cat qubit to data ones is exponentially suppressed by the cat size.
\label{fig:XX_measurement}}
\end{figure}
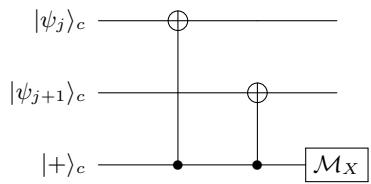
\begin{figure*}[t]
\includegraphics[width=0.8\linewidth]{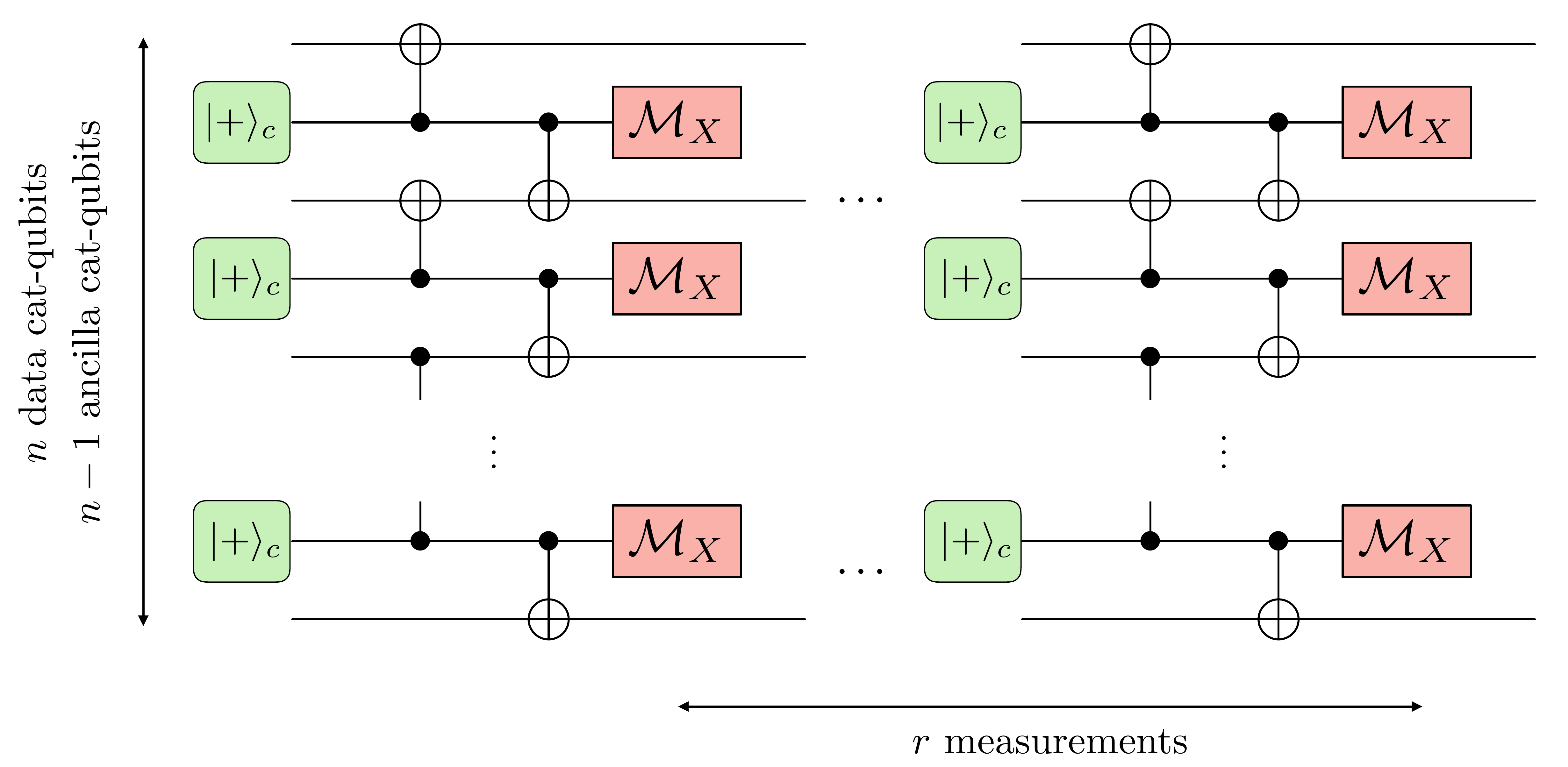}
\caption{
Quantum error correction circuit for a repetition cat qubit. One repetition cat qubit, composed of  data cat qubits and ancilla cat qubits (in green), is protected from errors by performing $r$ rounds of error detection. 
\label{QEC}}
\end{figure*}

In order to make these measurements fault tolerant, they are repeated $r$ times, where $r$ is optimized, taking into account the CNOT and measurement errors. We depict in Fig.~\ref{QEC} the full circuit for QEC on the repetition cat qubit. An optimal decoding, based on the outcome of the $(n-1)r$ ancilla measurement{s}, is then used to correct the effective phase-flip errors (see Ref.~\cite{Kelly-Nature-2015} for a practical implementation of such an optimal decoding). 

\paragraph*{Preparation of $\ket+_L$ and measurement of $X_L$.} 
 The  preparation of $\ket{\pm}_L = \ket{\pm}_c^{ \otimes n}$ can be performed transversally from the preparation of $\ket{\pm}_c$: $\mathcal{P}_{\ket{\pm}_L} = (\mathcal{P}_{\ket{\pm}_c})^{ \otimes n}$. A subsequent step of quantum error correction  enables us to reduce the preparation infidelity according to the distance of the repetition code. 
 
 Similarly, the measurement of only one cat qubit from the repetition cat qubit using $\mathcal{M}_X$ already implements a measurement of $X_L$. However, to ensure fault tolerance, $\mathcal{M}_{X_L}$ is implemented by measuring $\mathcal{M}_X$ on all the cat qubits and then making a majority vote.
 
 \paragraph*{Logical CNOT gate.}
This gate is not required in our set of universal gates and can be suppressed from the set $\mathcal{S}_L$. However, its implementation is easy and can lead to more economical circuits for the realization of certain algorithms. Indeed, the logical CNOT gate is simply obtained from the physical one by performing $n$ CNOT gates in a transversal manner, as depicted in Fig.~\ref{fig:Logical_CNOT}.
 \begin{figure}[h]
\includegraphics[width=.35\linewidth]{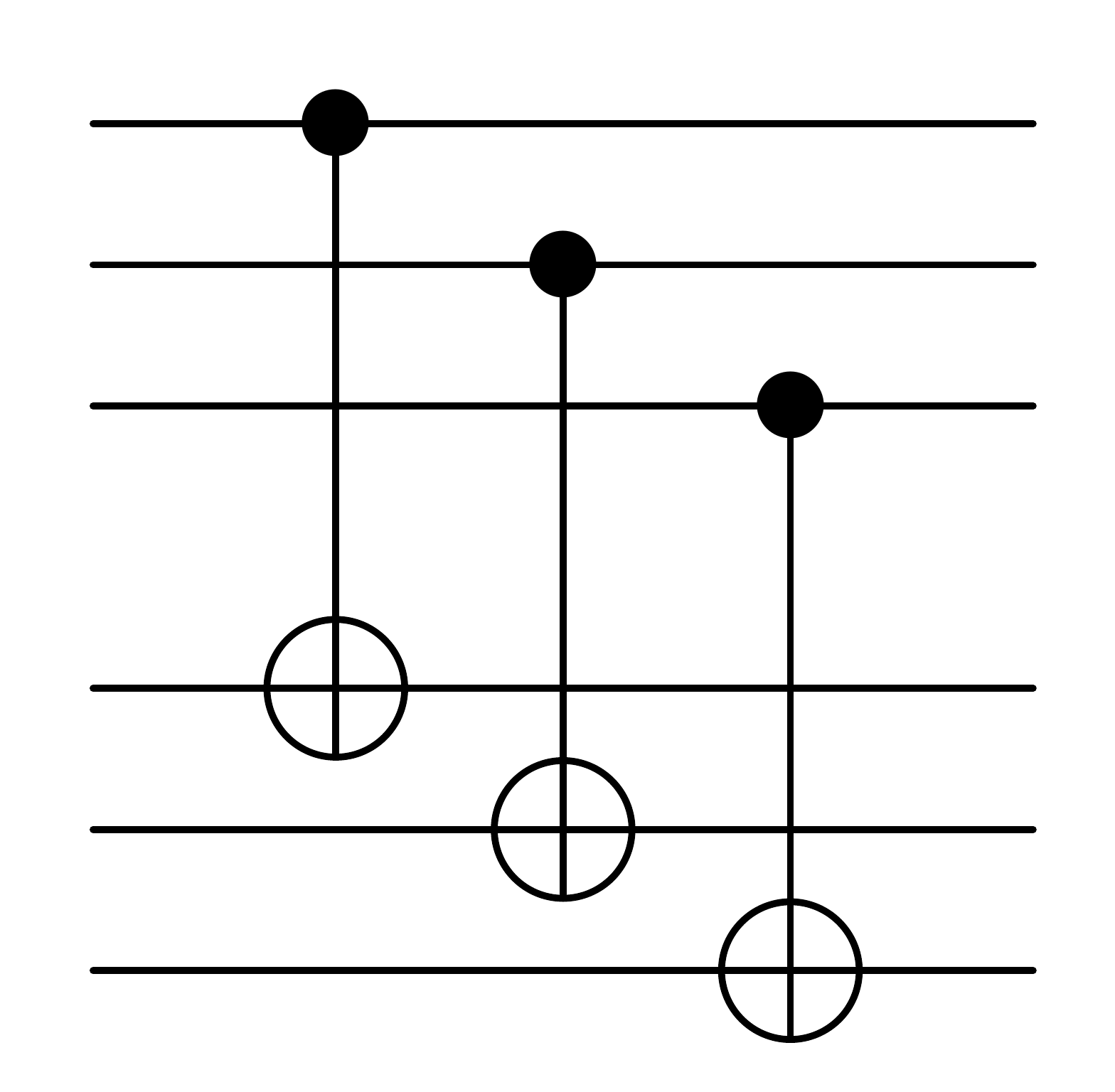}
\caption{\label{fig:Logical_CNOT}Transversal implementation of the logical CNOT. For clarity, the number of cat qubits per repetition cat qubit is $n=3$. Here the three upper lines represent the control repetition cat qubit and the three bottom ones represent the target one.}
 \end{figure}

The fault tolerance of the logical CNOT comes from the transversal construction of the circuit, which prevents the forward propagation of errors that could lead to an uncorrectable logical error. For instance, a phase-flip error occurring on the $k$th cat qubit of the logical target qubit $Z^{(k)}_2$ is converted through the circuit to a correlated $Z^{(k)}_1 Z^{(k)}_2$ error on two different blocks, but it cannot induce an error on a qubit $j \neq k$. Such errors are detected and corrected by the QEC stage performed after the Toffoli gate's execution. Furthermore, each CNOT gate at the level of cat qubits is bias preserving, and, therefore, the bit-flip errors remain exponentially suppressed at the repetition code level.

\paragraph*{Logical Toffoli gate.}
The transversal application of $n$ physical Toffoli gates in a similar fashion as Fig.~\ref{fig:Logical_CNOT} does not yield a Toffoli gate at the logical level. Instead, the logical Toffoli requires $n^2$ physical Toffoli gates, stacked as shown in Fig.~\ref{fig:Logical_Toffoli}. 
\begin{figure}[h!]
\includegraphics[width=\linewidth]{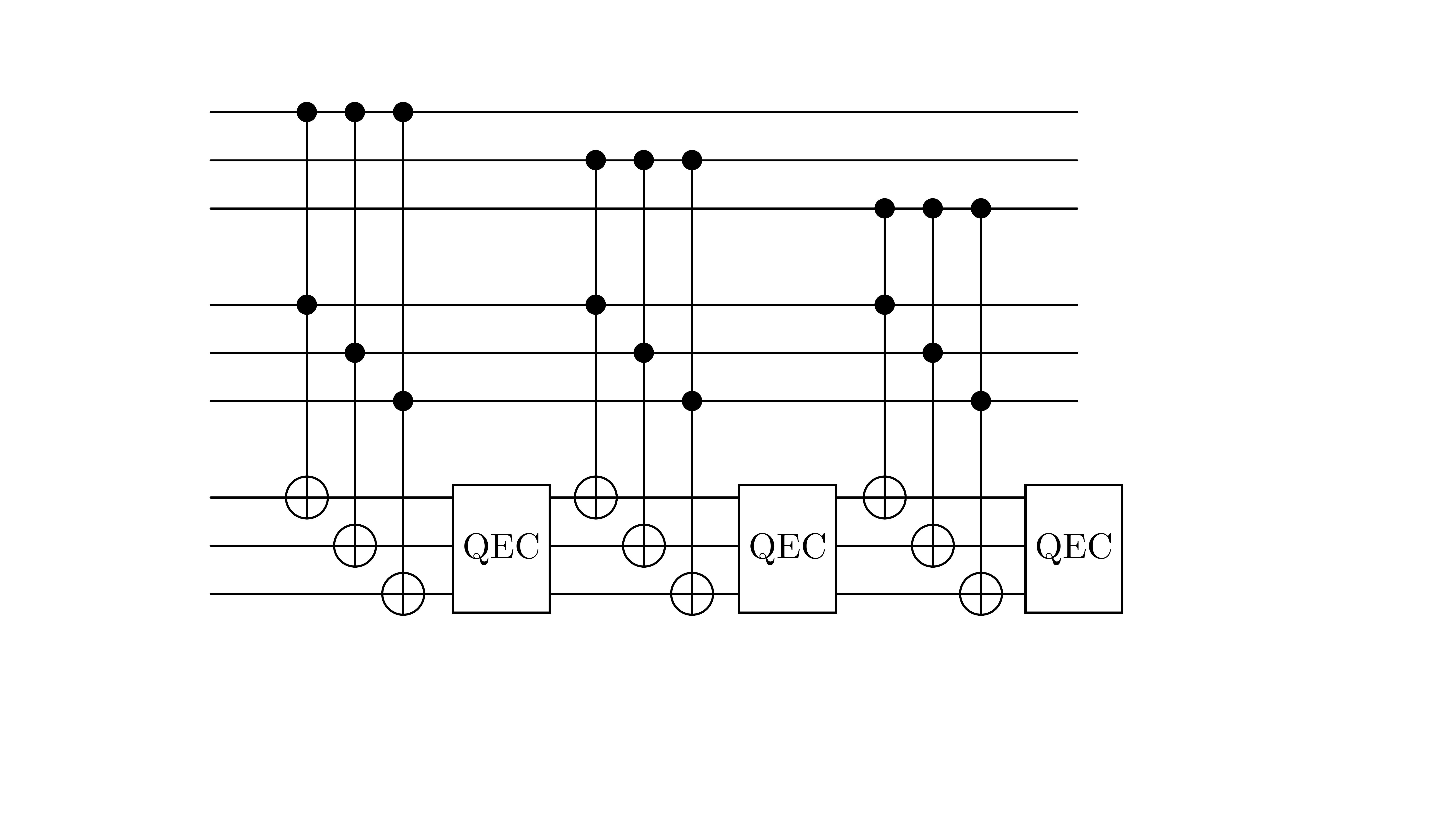}
\caption{\label{fig:Logical_Toffoli}Pieceable fault-tolerant Toffoli circuit. Each block of $n=3$ cat qubits represents one  repetition cat qubit.}
\end{figure} 
 
 The circuit can be understood as follows. Since the CNOT gate is transversal, the group of three physical Toffoli gates on the left-hand side of the circuit execute a logical CNOT between the logical qubits 2 and 3 (middle and lower blocks) conditioned on the first physical cat qubit of the first logical qubit being in the state $\ket 1_c$. Similarly, the group of three Toffoli gates in the middle (respectively, on the right) of the circuit also execute a CNOT on logical qubits 2 and 3 when the second (respectively, third) physical cat qubit of the first logical qubit is in the state $\ket 1_c$. Now, recall from Sec.~\ref{sec:bias} that the logical computational states of the first block are given by:
$ \ket0_L =  \tfrac{1}{(\sqrt2)^{n-1}} \sum \limits_{j \in \{0,1\}^n, |j|\ \text{even}}\ket j$ and $ \ket1_L =  \tfrac{1}{(\sqrt2)^{n-1}} \sum \limits_{j \in \{0,1\}^n, |j|\ \text{odd}}\ket j .$
\begin{figure*}[t!]
\includegraphics[width=.91\linewidth]{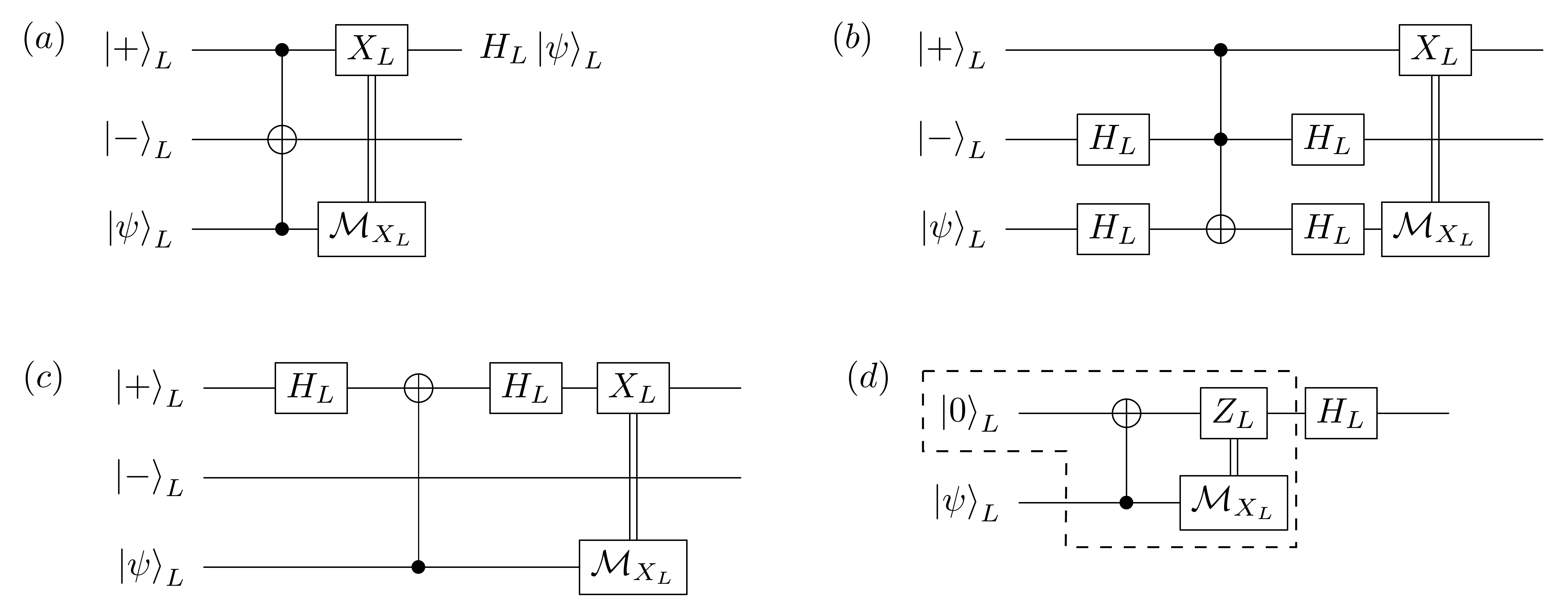}
\caption{\label{fig:Logical_Hadamard} Logical Hadamard circuit (a). The circuits (b,c,d) are equivalent circuits that help to understand why the circuit (a) implements a logical Hadamard gate. In circuit (b), the "CNOT" gate between the repetition cat qubits 2 and 3 is replaced by the equivalent circuit where the control and target roles are switched with the addition of Hadamard gates before and after the gate. It becomes clear in circuit (b) that the second repetition cat qubit plays no role: It is initialized in $\ket{-}_L$, transformed to $\ket{1}_L$ after the first Hadamard, thus always triggers the corresponding part of the control of the Toffoli, before being converted to $\ket{-}_L$ by the second Hadamard. Actually, the second repetition cat qubit is needed only because we do not readily have a controlled-phase gate available at the repetition cat qubit level (but here we show that it can be done with the Toffoli gate plus one ancilla repetition cat qubit). The idle role of the second repetition cat qubit is represented more simply in circuit (c), where again the "CNOT" between the repetition cat qubit 1 and 3 is replaced by its equivalent circuit where the control and target roles are exchanged with the addition of Hadamard gates. Finally, in circuit (d) (where we omit the second repetition cat qubit for clarity), the first Hadamard of the first line and the preparation of $\ket{+}_L$ are replaced by the equivalent preparation of $\ket{0}_L$ and the second Hadamard is commuted through the $X_L$, producing a $Z_L$ gate. The remaining circuit in the dashed box is just a teleportation of the state $\ket\psi_L$ of the second repetition cat qubit to the first one. After the state is teleported, the remaining Hadamard gate $H_L$ is applied, thus establishing the equivalence between circuit (a) and a logical Hadamard.}
\end{figure*}
The circuit works {in} this way: When the first logical qubit is in the $\ket0_L$ state, an even number of logical CNOTs is performed on the second and third logical qubits, which amounts to the identity operation. On the other hand, the input $\ket1_L$ state results in an odd number of logical CNOTs being performed, thus actually performing one logical CNOT. In other words, this circuit implements a logical "controlled-CNOT", that is, a Toffoli gate.
 
 We now discuss the fault tolerance of this circuit. First, as in the case of the CNOT gate, we note that each Toffoli gate at the level of the cat qubits is bias preserving, and, therefore, the bit-flip errors remain exponentially suppressed for the logical gate. Now, a phase-flip error occurring on any cat qubit of the first two logical qubit $Z_{1,2}$ (we omit the superscript for clarity) commutes with the successive physical Toffoli gates of the circuit:
 \[
 \text{Toffoli}\ Z_{1,2} = Z_{1,2}\ \text{Toffoli}.
 \]
 As a result, an error acting on any qubit of the first two logical qubits does not spread through the circuit and is corrected by the QEC stage performed after the gate's execution. Unfortunately, {phase-flip} errors on cat qubits of the third logical qubit do not commute with the Toffoli unitary and produce an extra error $U_{\text{err}} = \tfrac12(I_1+Z_1)  I_2 + \tfrac12(I_1-Z_1)  Z_2 $:
 \[
 \text{Toffoli}\ Z_3 = U_{\text{err}} Z_3 \text{Toffoli}
 \]
As the circuit is transversal for the second and third blocks of physical qubits (that is, the $k$th cat qubit of logical qubit 2 is only connected to the $k$th cat qubit of logical qubit 3), the forward propagation of an error from the third block to the second one cannot cause a logical error. On the other hand, because of the nontransversality of operations between the first and third logical qubits, a $Z_3$ error spreads on \textit{all} qubits of the first logical qubit, resulting in an uncorrectable logical $Z_L$ error. Though this result means that the full circuit is not fault tolerant, a logical error can arise only if a \textit{majority} of the qubits of the first logical qubit have been contaminated. To prevent this error, we perform two stages of QEC \textit{during} the execution of the circuit, after one-third and two-thirds of the full circuit are executed, as shown in Fig.~\ref{fig:Logical_Toffoli}. 

In this manner, any physical error  propagating from the third logical qubit is corrected before it spreads beyond the code distance. The exploited idea here is that, while a nontransversal circuit might not be fault-tolerant, pieces of the circuit could still be, thus, adding extra stages of error correction at carefully chosen locations in between these pieces can provide full fault tolerance. This idea is  introduced and studied in Ref.~\cite{Yoder2016}, and these circuits are called \textit{pieceable fault tolerant}.

Note that while the above analysis ensures that the errors do not propagate in an uncontrolled manner,  one also needs to prevent an accumulation of the nonpropagating errors (above the threshold of the repetition code) between two rounds of QEC. Indeed, Fig.~\ref{fig:Logical_Toffoli} represents the Toffoli circuit for a 3-qubit repetition code, and whenever we increase the code distance, we also need to add new blocks of QEC at appropriate places to avoid an accumulation of nonpropagating errors on the first repetition cat qubit. There are various ways to achieve this addition. The most straightforward one is inspired by the fact that a concatenation of repetition codes is still a repetition code.  More precisely, the measurements of the stabilizers of the lowest level repetition code are enough to reconstruct the value of the stabilizers at any higher level of encoding. One can, therefore, extend the logical Toffoli circuit of~Fig.~\ref{fig:Logical_Toffoli}  to a  $k^m$-qubit repetition code, where each data qubit meets at most $k$ gates between two rounds of QEC. A thorough numerical study and benchmarking of the accuracy threshold for such circuits is topic of current research and will appear in a forthcoming paper.

\paragraph*{Construction of Hadamard gate.}
In order to establish the universality of the set $\{\mathcal{P}_{\ket\pm_L}, \mathcal{M}_{X_L}, X_L, \text{Toffoli}\}$, we show in Fig.~\ref{fig:Logical_Hadamard} how a Hadamard gate can be built out of this set.  Its fault tolerance is trivially derived from the fault tolerance of each logical gate in the circuit.

We end this section by mentioning that while we have the possibility of correcting errors after the realization of each gate, it is enough to keep the record of errors by measuring the error syndromes and only apply an error correction before the Toffoli gates. Indeed, the Clifford operations map Pauli operators to Pauli operators and therefore it is only needed to keep track of the errors to update the Pauli frame in software~\cite{Knill2005}. On the other hand, the Toffoli gate is not a Clifford gate and does not map Pauli operators to Pauli operators. One therefore needs to perform error correction before the application of such a gate~\cite{Chamberland-Quantum-2018}. 

 \section{Error analysis}\label{sec:error_analysis}

We now study the performance of the physical gates of Sec.~\ref{sec:physical_gates} in a realistic setup.  We note that, while the bias preserving character of the state preparation $\cP_{\ket{\pm}_c}$ and the measurement $\cM_X$ is ensured by their definition, the error analysis  of our implementation of rotations around the $Z$ axis has been previously discussed in Ref.~\cite{Mirrahimi2014}. Therefore, we focus  our analysis to the case of $X$, CNOT and Toffoli gates. First, we analyze a model with two main sources of errors: the errors caused by single-photon loss in the cavity, which is the principal "physical" source of decoherence, and the errors induced by the finite gate time (nonadiabaticity). Then, we give numerical evidence that similar performance can be expected in the presence of other sources of errors, such as thermal excitations and dephasing.

In our implementation of the  $X$, CNOT and Toffoli gates, the loss of a single photon in the cavity changes the parity of the number of photons. As a result, the effect of photon loss on cat qubits is mostly to induce phase flips $\ket{+}_c \leftrightarrow \ket{-}_c$, but it rarely causes bit flips $\ket{0}_c \leftrightarrow \ket{1}_c$. Mathematically, this difference can be understood by looking at the matrix elements coupling these states, 
\begin{align*}
|\bra{\cC_{\alpha(t)}^+} \ba \ket{\cC_{\alpha(t)}^-}  |^2 &= |\alpha|^2 \tanh(|\alpha|^2) \underset{|\alpha|^2 \rightarrow +\infty}{\sim} |\alpha|^2  \\
|\bra{-\alpha(t)} \ba \ket{\alpha(t)} |^2 &= |\alpha|^2 e^{-2|\alpha|^2} \rightarrow 0
\end{align*}
If the only source of phase-flip errors is the photon loss, the noise bias, defined as the ratio of the phase-flip rate to the bit-flip rate $\eta = p_Z/ p_X$  scales as $\eta \sim e^{-2|\alpha|^2}$ for large values of $|\alpha|^2$. This situation is the case for the $X$ gate: As explained in Sec.~\ref{sec:physical_gates}, the phase-flip errors due to the adiabatic approximation disappear as soon as the  Hamiltonian{ $-\pi/T\ba^\dag\ba$} is added. Furthermore, the precision of the gate is ensured by the fact that the rotation of the cat states {leads} to a topological phase $\Cp\rightarrow \Cp$ and $\Cm \rightarrow -\Cm$. This phase is not affected by the imprecisions in the rotation angle. Indeed, the phase of the coherent states $\ket{\pm\alpha}$ is locked to the phase of the pump drives. In this sense, each cat qubit is defined with respect to its own pumps. Therefore, even if the rotation angle is not precisely $\pi$, which could happen, e.g., because of the amplitude and phase fluctuations of the pumping drive, the state still accumulates a topological $\pi$ phase with respect to its local oscillator. The same argument on the gate precision holds for the CNOT and Toffoli gates, and is discussed further in Sec. \ref{sec:experimental}. In the rest of this section, we will analyze the errors induced by  photon loss and nonadiabatic effects on the CNOT and Toffoli gates.

\begin{figure*}[t!]
\includegraphics[width=.91\linewidth]{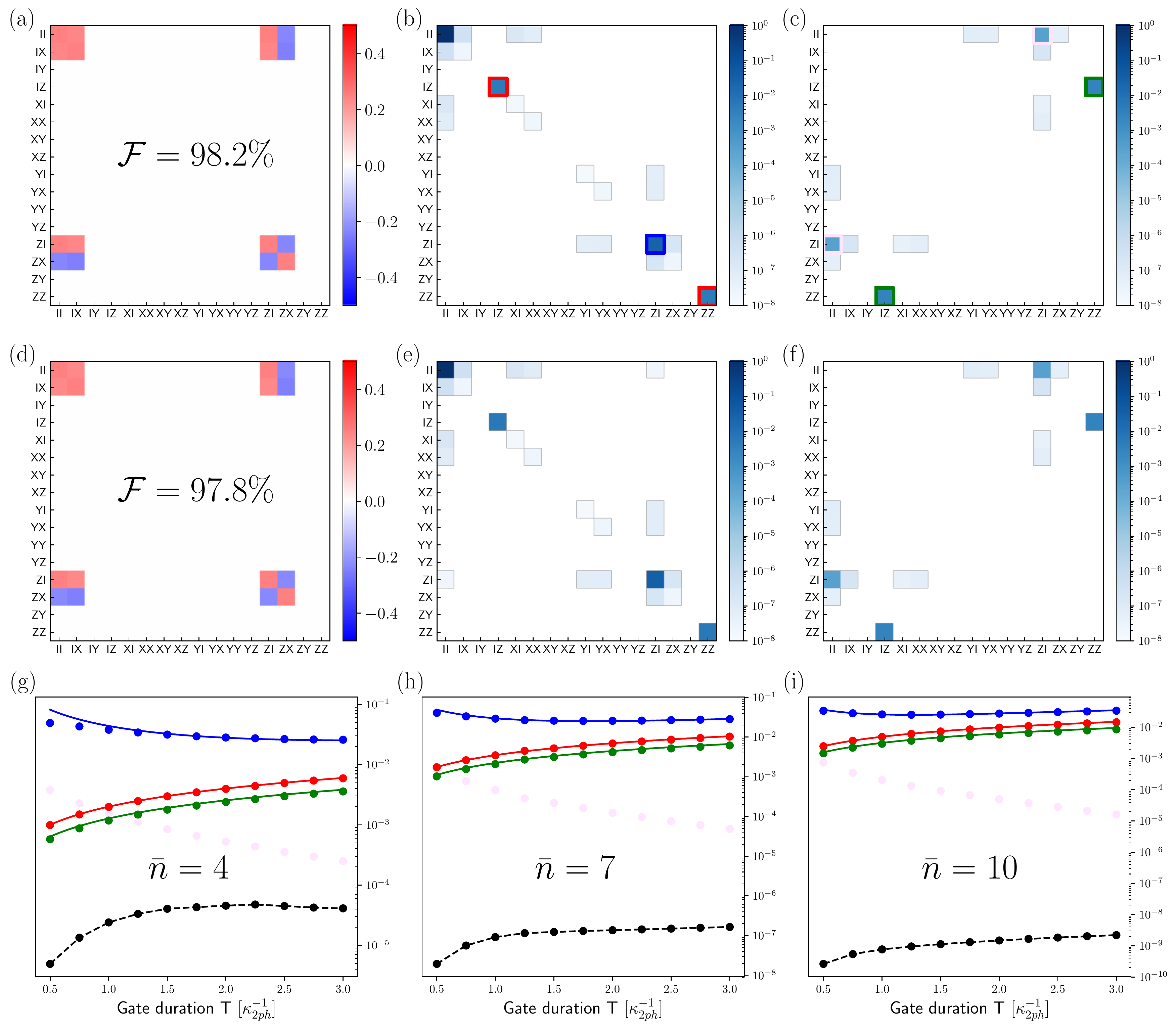}
\caption{\label{fig:CNOT} Process tomography of the CNOT gate in the presence of noise. The CNOT process presented in Sec.~\ref{sec:physical_gates} is numerically simulated for $\bar n = 7$ photon cat qubits using two different error models. First, we consider photon loss on both modes $\kappa_{1\text{ph}}\mathcal{D}[\ba] + \kappa_{1\text{ph}}\mathcal{D}[\hat b]$ (a-c). Then, we consider a more elaborate error model including photon loss $\kappa_{1\text{ph}}(1+n_{\text{th}})\mathcal{D}[\ba] + \kappa_{1\text{ph}}(1+n_{\text{th}})\mathcal{D}[\hat b]$, thermal excitations $\kappa_{1\text{ph}} n_{\text{th}} \mathcal{D}[\ba^\dagger] + \kappa_{1\text{ph}} n_{\text{th}} \mathcal{D}[\hat{b}^\dagger]$ ($n_{\text{th}} = 10\%$) and dephasing on both modes $\kappa_{\phi}\mathcal{D}[\ba^\dagger\ba] + \kappa_{\phi}\mathcal{D}[\hat{b}^\dagger\hat{b}]$ (d-f). 
In both cases, we set $\kappa_{1\text{ph}}/\kappa_{2\text{ph}} = 10^{-3}$ and the gate time is chosen optimal $T^{*} = [2\bar n \sqrt{\pi \kappa_{1\text{ph}}\kappa_{2\text{ph}}}]^{-1} \approx 1.27 \kappa_{2\text{ph}}^{-1}$ (see the main text). 
We plot the real part of the process matrix $\chi$ (a,d), and the real (b,e) and imaginary (c,f) part of the error matrix $\chi^{\text{err}}$. In the lower row (g,h,i), we check the validity of our theoretical error model for photon loss for various gate times and cat sizes. While the dots illustrate the simulation results where the full master equation in the presence of loss are considered, the plain lines correspond to the analytical formula provided in the main text. The blue dots correspond to the diagonal process matrix element corresponding to the $Z_1$, the red dots correspond to the coinciding diagonal matrix elements  corresponding to  $Z_2$ and $Z_1Z_2$. The green dots correspond to the off-diagonal elements corresponding to the coherence between $Z_2$ and $Z_1Z_2$ errors. The pale magenta dots correspond to the off-diagonal elements corresponding to coherence between $Z_1$ and $I$, this coherence is due to high-order nonadiabatic effects and is not included in our theory. The black dots correspond to all remaining errors, including bit flip-type ones. It is clear that such errors remain exponentially suppressed. 
}
\end{figure*}

\paragraph*{CNOT gate corrupted by photon loss.}
In order to understand the effect of a loss of photon during the execution of the CNOT,  let us consider the unitary operation approximately generated by the two dissipation channels $\mathcal{L}_{\ba}$,  $\mathcal{L}_{\hat{b}}(t)$ of Sec.~\ref{sec:physical_gates}. Indeed, in the presence of the above driven dissipations where the dynamics is confined in the cat qubit subspace, we approximately implement a unitary operation of the form:
\[
\mathcal{U}(t) = \ket\alpha\bra\alpha \otimes I + \ket{-\alpha}\bra{-\alpha} \otimes e^{i \frac\pi T t \hat{b}^\dagger \hat{b}}
\]
with $\mathcal{U}(0) = I \otimes I$ and $\mathcal{U}(T) = \text{CNOT}$. 

First, we investigate the effect of a loss of a photon of the control mode $\ba$. Denote by $\mathcal{E}_{\ba}$ the noisy quantum operation performed instead of the CNOT when the control mode is lossy, and by $t \in [0,T]$ the random time at which such a loss occurs. We have:
\begin{align*}
\mathcal{E}_{\ba} &= \mathcal{U}(T - t) [\hat{a} \otimes I] \mathcal{U}(t) \\
&=\alpha \ket\alpha\bra\alpha \otimes I - \alpha \ket{-\alpha}\bra{-\alpha} \otimes e^{i \pi \hat{b}^\dagger \hat{b}} \\
&= [ \hat{a} \otimes I ] \text{CNOT}
\end{align*}
which can be written in terms of Pauli operators for the cat qubits as 
\[
\mathcal{E}_{\ba} = Z_1 \text{CNOT}. 
\]
In other words, a photon loss on the control cat qubit causes a phase-flip on that qubit but does not affect the target cat qubit.

On the other hand, a photon loss occurring on the target cat qubit $\hat b$ at time $t$ propagates as
\begin{align*}
\mathcal{U}&(T - t)  [I \otimes \hat{b}] \mathcal{U}(t) =\\ &(I \otimes \hat{b})(\ket\alpha\bra\alpha \otimes I 
+ e^{-i\pi\frac{T-t}{T}} \ket{-\alpha}\bra{-\alpha} \otimes e^{i \pi \hat{b}^\dagger \hat{b}})=\\
& (I \otimes \hat{b})(\ket\alpha\bra\alpha \otimes I + e^{-i\pi\frac{T-t}{T}} \ket{-\alpha}\bra{-\alpha} \otimes I) \text{CNOT}.
\end{align*}
The resulting error 
\[
I \otimes \hat{b} (\ket\alpha\bra\alpha \otimes I + e^{-i\pi\frac{T-t}{T}} \ket{-\alpha}\bra{-\alpha} \otimes I)
\]
induced by the propagation of the photon loss can be expressed in terms of the Pauli operators of cat qubits as 
\[
\hat{U}_{\text{err}}(\theta) = \frac 12 (1+Z_1) Z_2 + \frac 12 e^{i\theta}(1-Z_1)Z_2
\]
where $\theta = -i \pi (1 - t/T)$ is a random phase. The time of the jump being uniformly distributed over the interval $[0,T]$, the noisy operation $\mathcal{E}_{\hat b}$ can be written
\begin{align}\label{eq:coh}
\mathcal{E}_{\hat b}(\rho) &= \bar n \kappa_{1\text{ph}} T \int_{-\pi}^0 \frac{d\theta}{\pi} \hat{U}_{\text{err}}(\theta) \tilde\rho \hat{U}_{\text{err}}(\theta)^\dagger \notag\\
&=\bar n \kappa_{1\text{ph}} T [ \frac 12 Z_2 \tilde\rho Z_2 + \frac 12 Z_1Z_2 \tilde\rho Z_1Z_2\notag\\
&\hspace{1,5cm} + \frac i\pi Z_1Z_2 \tilde\rho Z_2 -\frac i\pi Z_2 \tilde\rho Z_1Z_2 ]
\end{align}
where $\tilde\rho = \text{CNOT} \rho \text{CNOT}$ is the image of $\rho$ by a perfect CNOT operation and $\bar n \kappa_{1\text{ph}} T$ is the average number of photons lost in each mode during the gate's execution. Written in this form, it is clear that the operation $\mathcal{E}_{\hat b}$ is a perfect CNOT gate followed by some noise given by the operators appearing in the equation above. The first two terms indicate that the effect of photon loss on the target cat qubit produces two types of error of the same strength: phase flips on the target cat qubit $\frac 12 Z_2 \tilde\rho Z_2$ as well as a correlated phase flips on both qubits $\frac 12 Z_1Z_2 \tilde\rho Z_1Z_2$, with some degree of coherence between these two errors. 

When losses on both modes are taken into account, the noisy CNOT $\mathcal{E}_{\ba, \hat b}$ can be expressed as a Kraus sum:
\[
\mathcal{E}_{\ba, \hat b}(\rho) =  \sum \limits_{k = 1,2,3} \hat M_k \tilde\rho \hat M_k^\dagger 
\]
where ($r = \tfrac 12 \arcsin(2/\pi)$):
\begin{align*}
\hat M_1 &= \sqrt{\bar n \kappa_{1\text{ph}} T} Z_1\\
\hat M_2 &= \sqrt{\tfrac{\bar n \kappa_{1\text{ph}} T}{2}}(\cos r I_1 + i \sin r Z_1) Z_2 \\
\hat M_3 &= \sqrt{\tfrac{\bar n \kappa_{1\text{ph}} T}{2}}(\sin r I_1 + i \cos r Z_1) Z_2.
\end{align*}

\paragraph*{Errors induced by nonadiabaticity.}
The Hamiltonian 
\[
\hat H = \frac12 \frac \pi T \frac{ \hat a -\alpha}{2\alpha} \otimes (\hat{b}^\dag \hat b - \bar n) + h.c.
\]
compensates most of the errors induced by the finite gate time $T$. Using the adiabatic elimination techniques of Ref.~\cite{Azouit2016}, it is possible to characterize the remaining error and show that it is composed only of phase flips on the control cat qubit $Z_1$, with a rate proportional to $(\bar n \kappa_{2\text{ph}} T^2)^{-1}$. The exact coefficient of proportionality can be estimated by a numerical fit and is well approximated by $2\pi$, giving the phase-flip  probability: 
\[
p_{Z_1}[\text{nonadiabaticity}] = (2 \pi \bar n \kappa_{2\text{ph}} T)^{-1}.
\]
A more thorough study of the errors induced by the approximate Hamiltonian will be given in a forthcoming paper.

\paragraph*{{Numerical simulations.}}
In order to check the validity of this  error model, we perform a numerical process tomography of the CNOT gate [Figs.~\ref{fig:CNOT}(a,b,c)]. This process is done by simulating the full master equation of the system in the presence of photon loss. The process matrix $\chi$ plotted in Fig.~\ref{fig:CNOT}(a) completely characterizes the quantum operation $\mathcal{E}$ performed via the relation 
\[
\mathcal{E}(\rho) = \sum \limits_{mn} \chi_{mn} P_m \rho P_n^\dagger
\]
where $\{P_j\}$ is the set of two-qubit Pauli operators. The gate fidelity $\mathcal{F}$ is defined as~\cite{Nielsen2011}
\begin{equation}\label{eq:def_fid}
\mathcal{F}(U,\cE) = \min_{\ket{\psi}}\mathcal{F}(U\ket{\psi},\cE(\ket{\psi}\bra{\psi})) 
\end{equation}
where {$U=\text{CNOT}$} is the perfect CNOT operation and the {right-hand side} represents the minimum over all two-qubit state fidelities. The unitary of the perfect CNOT is factored out in order to obtain the process error matrix $\chi^{\text{err}}$ (real part in (b), imaginary part in (c)), which characterizes the noise alone: 
\[
\mathcal{E}(\rho) = \sum \limits_{mn} \chi^{\text{err}}_{mn} P_m \tilde{\rho} P_n^\dagger
\] 
with $\tilde{\rho} = \text{CNOT} \rho \text{CNOT}$  the image of $\rho$ by a perfect CNOT. In other words, we decompose the noisy CNOT into a perfect CNOT followed by some noise process, characterized by the process error matrix $\chi^{\text{err}}$. As  can be seen in the real part of $\chi^{\text{err}}$ (Figure~\ref{fig:CNOT}-b), photon loss and nonadiabaticity cause only phase-flip errors $Z_1$, $Z_2$ or $Z_1Z_2$.

We further investigate our theoretical model for errors caused by photon loss by plotting (Fig.~\ref{fig:CNOT}(g,h,i)) the values of the coefficients of the error matrix $\chi^{\text{err}}$ (marked by colored squares) as a function of the gate duration. Here, the blue dots correspond to phase-flip errors on the control cat qubit $Z_1$ induced by a combination of nonadiabatic errors and the photon loss. The plain blue line corresponds to our analytical formula
$$
p_{Z_1}=\bar n \kappa_{1\text{ph}}T+(2\pi\bar n\kappa_{2\text{ph}}T)^{-1},
$$
which is found through the analysis above together with second-order perturbation techniques following Ref.~\cite{Azouit2016}. The red dots represent the phase-flip errors on target qubit $Z_2$ and the correlated phase-flip errors  $Z_1 Z_2$ (red dots). These values coincide, and, following our analysis, they are given by
$$
P_{Z_2}=P_{Z_1Z_2}=\bar n \kappa_{1\text{ph}}T/2.
$$ 
This result is represented by the plain line in red. The off-diagonal term representing the coherence between $Z_2$ and $Z_1Z_2$ errors (green dots) also fits very well our expectation  from  Eq.~\eqref{eq:coh}. The pale purple dots correspond to the off-diagonal term representing the coherence between $I$ and $Z_1$ errors. In order to capture such a coherence, one needs to push the nonadiabatic perturbation techniques~\cite{Azouit2016} up to third order, which will be subject of future work. Most importantly, the remaining errors (namely, the ones that contain an X or Y Pauli operator) represented by the black lines are exponentially suppressed by the cat size. This result proves the bias preserving aspect of the gate.

\paragraph*{Gate fidelity and optimal gate time.}
The phase-flip errors occurring on the control mode are caused by two sources that are very different in essence: When the gate time is increased, the "environment"-induced errors, caused by photon loss, are also increased, whereas phase-flip errors caused by nonadiabaticity are reduced. This opposite behavior gives rise to a finite optimal gate time $T^*$ for which the gate infidelity is minimal.

More precisely, taking into account {phase-flip} errors caused by photon loss and by non-adiabicity, the total {phase-flip} error probability on the control cat qubit is given by
\begin{align*}
p_{Z_1} &= p_{Z_1}[\text{photon loss}] + p_{Z_1}[\text{nonadiabaticity}] \\
&= \bar n \kappa_{1\text{ph}} T + (2 \pi \bar n \kappa_{2\text{ph}} T)^{-1}.
\end{align*}
The gate fidelity $\mathcal{F}$ of the implemented CNOT operation, defined by Eq.~\eqref{eq:def_fid}, is given by
\begin{align*}
\mathcal{F} &= \sqrt{1 - (p_{Z_1} + p_{Z_2} + p_{Z_1 Z_2})}\\
&= \sqrt{1 - 2 \bar n \kappa_{1\text{ph}} T -  (2 \pi \bar n \kappa_{2\text{ph}} T)^{-1}}.
\end{align*}

The highest value of the fidelity that can be achieved is set by the ratio $\kappa_{1\text{ph}} / \kappa_{2\text{ph}}$
\[
\mathcal{F} = \sqrt{1 - \sqrt{\frac 4 \pi \frac{\kappa_{1\text{ph}}}{\kappa_{2\text{ph}}} }},
\]
achieved for the optimal gate time 
\[
T^{*} = [2\bar n \sqrt{\pi\frac{\kappa_{1\text{ph}}}{\kappa_{2\text{ph}}}}]^{-1} \kappa_{2\text{ph}}^{-1}.
\]
{For the ratio $\frac{\kappa_{1\text{ph}}}{\kappa_{2\text{ph}}} = 10^{-3}$ considered in Fig.~\ref{fig:CNOT}, this theoretical formula predicts a gate fidelity of $\mathcal{F} = 98.2 \%$, in agreement with the numerical simulation.}

\paragraph*{Addition of new noise processes.}
As discussed in Ref.~\cite{cohen-thesis-2017}, in the presence of the two-photon pumping scheme, any physical noise process with a local effect in the phase space of the harmonic oscillator causes bit flips that are exponentially suppressed in the size of the cat qubits, thus preserving the biased structure of the noise. We now provide a numerical evidence of this fact for a more elaborate set of physical noise processes for the superconducting cavity: photon loss $\ba$ , thermal excitation $\ba^\dagger$ with a non-zero temperature, and  photon dephasing $\ba^\dagger \ba$.

In Fig.~\ref{fig:CNOT}, we characterize the performed operation by plotting the process matrix $\chi$ (d), and the real part (e) and imaginary part (f) of the error matrix. In this simulation, $\kappa_{1\text{ph}}/\kappa_{2\text{ph}} = 10^{-3}$, the photon loss is given by $\kappa_{1\text{ph}}(1+n_{\text{th}})\mathcal{D}[\ba]$ and thermal excitations by $\kappa_{1\text{ph}} n_{\text{th}} \mathcal{D}[\ba^\dagger]$ with $n_{\text{th}} = 10\%$, and the dephasing on the cavity is given by $\kappa_{\phi}\mathcal{D}[\ba^\dagger\ba]$ with $\kappa_{\phi} = \kappa_{1\text{ph}}$.

Note that the resulting error matrix and gate fidelity are barely affected by the added thermal excitations and photon dephasing. The addition of thermal noise and dephasing slightly decrease the fidelity of the operation, from $98.2\%$ to $97.8\%$, but as expected, this decrease is caused by an increased rate of phase-flip errors, while all bit-flip errors remain exponentially suppressed.

\paragraph*{Numerical considerations.}
The numerical study of the CNOT gate in the presence of noise (Fig.~\ref{fig:CNOT}) requires the simulation over a time $T$ of the Lindblad master equation 
\[
\begin{split}
\dot{\rho} = -i [\hat H, \rho] + \kappa_{2\text{ph}} \mathcal{D}[\hat L_{\hat a}] + \kappa_{2\text{ph}} \mathcal{D}[\hat L_{\hat b}(t)] \\
+ \kappa_{1\text{ph}} \mathcal{D}[\hat a]  + \kappa_{1\text{ph}} \mathcal{D}[\hat b] 
\end{split}
\]
for 16 different initial states, where $\hat H, \hat L_{\hat a}$ and $\hat L_{\hat b}(t)$ are defined in Sec.~\ref{sec:physical_gates}. The numerical computations were performed in parallel using the cluster of Inria Paris, composed of 68 nodes for a total of 1244 cores. The nodes are divided in a few hardware generations: 28 biprocessors Intel Xeon X5650 of 6 cores, 12 biprocessors E5-2650v4 2.20 of 12 cores, 16 biprocessors XeonE5-2670 of 10 cores, 8 biprocessors E5-2695 v4 of 18 cores, 4 biprocessors E5-2695 v3 of 14 cores.
The simulation of the CNOT gate for cat qubits of $\bar n = 10$ and a gate time of $T = 3\kappa_{2\text{ph}}^{-1}$ takes about 13 h on the cluster. The simulation of the Toffoli gate with the same parameters would be about 2000 times longer, for this reason, we do not provide numerical simulations for the Toffoli gate in this paper. However, the analytical discussion that follows explains, in a similar manner to the CNOT gate (plain lines in Figs.~\ref{fig:CNOT}(g,h,i)), the expected error mechanisms and rates.

\paragraph*{{Toffoli gate corrupted by photon loss.}}
The effect induced by photon loss during the execution of the Toffoli gate can be derived in the same way as for the CNOT. A photon loss occurring on one of the two control modes $\ba$, $\hat b$ does not propagate to the other modes and results in a dephasing error $Z_1$ and $Z_2$, respectively. When the target mode $\hat c$ loses a photon, it gives rise to a correlated error between the three modes. More precisely, the noisy Toffoli operation $\mathcal{E}_{\ba, \hat b, \hat c}$ can be decomposed into a perfect Toffoli operation, again denoted by 
\[
\tilde \rho = \text{Toffoli}\ \rho\ \text{Toffoli}
\]
followed by a noise process expressed in Kraus form as
\[
\mathcal{E}_{\ba, \hat b, \hat c}(\rho) = \sum \limits_{k = 1,2,3,4} \hat M_k \tilde\rho \hat M_k^\dagger 
\]
\begin{align*}
\hat M_1 &= { \sqrt{\bar n \kappa_{1\text{ph}} T}} Z_1\\
\hat M_2 &= { \sqrt{\bar n \kappa_{1\text{ph}} T}} Z_2 \\
\hat M_3 &= { \sqrt{\bar n \kappa_{1\text{ph}} T}} [\cos r (I_1I_2 - \cZ_{12}) - i \sin r \cZ_{12}] Z_3 \\
\hat M_4 &= { \sqrt{\bar n \kappa_{1\text{ph}} T}}[\sin r (I_1I_2 - \cZ_{12}) - i \cos r \cZ_{12}] Z_3.
\end{align*}

where $\cZ_{12}= \frac 14 (I_1I_2 - Z_1 - Z_2 - Z_1Z_2)$ acts on the two control cat qubits.

\paragraph*{{Nonadiabatic effects on Toffoli implementation.}}
Because of the analogies in the way the CNOT and the Toffoli gates are implemented, it is useful to think of the Toffoli gate as a CNOT where the control state $\ket{-\alpha}$ is replaced by $\ket{-\alpha, -\alpha}$. In particular, the methods of~\cite{Azouit2016} that we use to characterize the effect of nonadiabaticity predict similar results for the Toffoli gate. The analysis of the errors induced by the approximate Hamiltonians is not the subject of this paper. However, we anticipate that the effect of the finite gate time is to dephase the ``trigger'' state $\ket{-\alpha, -\alpha}$ with respect to the other three possible states of the pair of control cat qubits. In terms of Pauli operator, this effect results only in phase-flip errors $Z_1$, $Z_2$ and $Z_1 Z_2$ on the two control cat qubits with equal probability $p = (4\pi\bar n \kappa_{2\text{ph}} T)^{-1}$ but it does not cause any error on the target cat qubit, or bit-flip-type errors. 

\paragraph*{Optimal gate time and gate fidelity.}
Taking into account phase-flip errors caused by photon loss and nonadiabaticity, the gate fidelity $\mathcal{F}$ of the implemented Toffoli operation is given by
\begin{align*}
\mathcal{F} &= [ 1 - p_{Z_1} - p_{Z_2} - p_{Z_3} \\
 &\hspace{1cm}- p_{Z_1 Z_2} - p_{Z_1 Z_3} - p_{Z_2 Z_3} - p_{Z_1 Z_2 Z_3}]^{\frac12} \\
 &= [1 - \frac{3}{4\bar n \pi \kappa_{2\text{ph}}T} -3\bar n \kappa_{1\text{ph}} T]^{\frac12}.
 \end{align*}
 
 This fidelity is maximum for the same gate time $T^* =  [2\bar n \sqrt{\pi\frac{\kappa_{1\text{ph}}}{\kappa_{2\text{ph}}}}]^{-1} \kappa_{2\text{ph}}^{-1}$ optimizing the CNOT gate, producing a gate fidelity of 
\[
\mathcal{F} = \sqrt{1 - \sqrt{\frac 9 \pi \frac{\kappa_{1\text{ph}}}{\kappa_{2\text{ph}}} }}.
\]

The ratio $\frac{\kappa_{1\text{ph}}}{\kappa_{2\text{ph}}} = 10^{-3}$ considered in Fig.~\ref{fig:CNOT} corresponds to a gate fidelity $\mathcal{F} = 97.3 \%$ with respect to a perfect Toffoli.
Note that the optimal gate time for the CNOT and the Toffoli gate decreases with the mean number of photons $\bar n$. 

 \section{toward experimental implementation}\label{sec:experimental}

\bgroup
\def\arraystretch{2.5}
 \begin{table*}[t]
 \begin{center}
    \begin{tabular}{| >{\centering\arraybackslash}p{3cm}  | >{\centering\arraybackslash}p{9cm} | >{\centering\arraybackslash}p{3.7cm} | >{\centering\arraybackslash}p{1.8 cm}  |}
    \hline
    Bias-preserving gates for cat qubits& Dissipation operators  & Hamiltonians & Experiments \\ \hline
   Identity and $\mathcal{P}_{\ket+_c}$& $\kappa_{2\text{ph}}\cD[\ba^2-\alpha^2]$ (mandatory) & None & \cite{Leghtas2015,Touzard-PRX-2018,lescanne-dephasing-2019} \\ \hline
   $\cM_X$ & None & $-\chi\ket{e}\bra{e}\ba^\dag\ba$ (mandatory) &\cite{Sun2014,Ofek-Petrenko-Nature-2016} \\ \hline
   $Z(\theta)$ and $\mathcal{P}_{\ket-_c}$& $\kappa_{2\text{ph}}\cD[\ba^2-\alpha^2]$ (mandatory) & $\epsilon_Z(\ba+\ba^\dag)$ (mandatory) &\cite{Touzard-PRX-2018} \\ \hline
   $X$ & $\kappa_{2\text{ph}}\cD[\ba^2-e^{\frac{2i\pi t}{T}}\alpha^2]$ (mandatory) & $-\frac{\pi}{T} \ba^\dag\ba$ (optional) & None \\ \hline
   CNOT & $\kappa_{2\text{ph}}\cD[\ba^2-\alpha^2]$ and $\kappa_{2\text{ph}}\cD[\bb^2-\alpha^2-\frac{\alpha}{2}(1-e^{\frac{2i\pi t}{T}})(\ba-\alpha)]$ (mandatory) & $\frac12 \frac \pi T \frac{ \hat a -\alpha}{2\alpha} \otimes (\hat{b}^\dag \hat b-|\alpha|^2) +$ h.c. (optional) &None \\ \hline
   Toffoli & $\kappa_{2\text{ph}}\cD[\ba^2-\alpha^2]$, $\kappa_{2\text{ph}}\cD[\bb^2-\alpha^2]$, and $\kappa_{2\text{ph}}\cD[\hat{c}^2 -\alpha^2+\frac{1}{4}(1-e^{\frac{2i\pi t}{T}})(\ba\bb-\alpha(\ba+\bb)+\alpha^2)]$ (mandatory) & $-\frac12 \frac \pi T \frac{\hat a-\alpha}{2\alpha}\otimes \frac{\hat b-\alpha}{2\alpha}\otimes (\hat{c}^\dag \hat c-|\alpha|^2) +$ h.c. (optional) &None \\ \hline
   $CZ(\theta)$ (optional) & $\kappa_{2\text{ph}}\cD[\ba^2-\alpha^2]$ and $\kappa_{2\text{ph}}\cD[\bb^2-\alpha^2]$ & $\epsilon_{ZZ}(\ba\bb^\dag+\bb\ba^\dag)$ &None \\ \hline
    \end{tabular}
  \caption{\label{table:exp} Dissipation operators and Hamiltonians required for universal and fault-tolerant quantum computation with cat qubits. The mandatory Hamiltonians and dissipators are required to achieve bias-preserving operations at the level of cat qubits which, embedded in a repetition code, lead to a universal set of fault-tolerant logical gates. The optional Hamiltonians reduce the phase-flip error rate induced by the {nonadiabatic effects} of the bias-preserving operations. Such an improvement can lead to drastic reduction of the number of the required cat qubits in a repetition code, in order to reach a certain desired error level. While some of the operations in the table have already been implemented in experiments with superconducting circuits, all the other ones should be simple modifications or extensions of   the current experiments. }
  \end{center}
 \end{table*}
\egroup

The ideas behind our proposal could be applied to  qubits possessing an extra degree of freedom, for which one type of error (bit flips or phase flips) is suppressed as soon as  the noise processes are local in this degree of freedom (here the locality is in the phase space, but one can also consider the locality in the actual space as in Majorana fermions). The pumped (driven-dissipative or Kerr-type) cat qubits are prototypes of such qubits where the information is encoded and stabilized in a non-local manner in the phase space of harmonic oscillators. The stabilization of cat qubits can, for instance, be implemented in the context of the vibrational degree of freedom of an ion, by engineering two-phonon dissipation and drive~\cite{Poyatos-PRL-96}. Throughout the past years, this idea has attracted a lot of attention in the context of superconducting circuits where such a two-photon driven dissipation can be systematically achieved using the four-wave mixing property of Josephson junctions and applying parametric  methods~\cite{Mirrahimi2014,Leghtas2015,Touzard-PRX-2018,lescanne-dephasing-2019} (see also Refs.~\cite{Puri-Blais-2017,Wang-Safavi-Naeini-2019,grimm-frattini-dephasing-2019} for the nondissipative approach known as the Kerr cat). We therefore pursue this trend by proposing a possible implementation of various dissipation and Hamiltonian operators involved in our scheme using parametric methods for superconducting circuits. 

We summarize in Table~\ref{table:exp} the dissipation operators and Hamiltonians required for the implementation of various bias preserving gates at the level of cat qubits. We note that, in this table some of the dissipation operators and Hamiltonians  are crucial as their realization enables us to achieve bias preserving gates where the bit flips remain exponentially suppressed with the cat size. Some other  Hamiltonians  are  optional and  enable us to reduce the phase-flip-type errors due to {nonadiabatic effects}. The implementation of the optional Hamiltonians can  put the phase-flip {error} probability of cat qubits well below the accuracy threshold of the repetition code and therefore reduce the number of required cat qubits in this code.  

First, we note that the two-photon driven dissipation required to stabilize the cat qubits is realized by parametrically coupling a high-Q superconducting cavity mode $\ba$ (frequency $\omega_a$), called the storage mode, to a low-Q one $\bd$ (frequency $\omega_d$), called the dump mode~\cite{Leghtas2015,Touzard-PRX-2018,lescanne-dephasing-2019}. Here, we recall the approach in these implementations. Coupling the two modes  $\ba$ and $\bd$ via a nonlinear element (a Josephson junction), and driving the system at frequency $2\omega_a-\omega_d$, one can engineer a  nonlinear interaction of the form
$$
H_{2\text{ph}}=(g_{2\text{ph}}\ba^2\bd^\dag+g_{2\text{ph}}^*\ba^{2\dag}\bd).
$$
In particular, the amplitude and phase of the coupling, given by the complex value $g_{2\text{ph}}$, is modulated by the pump (drive at frequency  $2\omega_a-\omega_d$) amplitude and phase and one can turn off such a coupling simply by turning off the pump. One can additionally, consider a resonant drive at frequency $\omega_d$, modeled by the Hamiltonian $H_d=\epsilon_d\bd^\dag+\epsilon_d^*\bd$. Therefore, the total system follows the master equation
$$
\frac{d}{dt}\rho=-i[g_{2\text{ph}}\ba^2\bd^\dag+g_{2\text{ph}}^*\ba^{2\dag}\bd,\rho]-i[\epsilon_d\bd^\dag+\epsilon_d^*\bd,\rho]+\kappa_d\cD[\bd]\rho.
$$
In this master equation, the dump mode $\bd$ mediates a two-photon exchange between the storage mode $\ba$ and its bath. More precisely, by adiabatically eliminating the dump mode (assuming $\kappa_d>|g_{2\text{ph}}|$ and $|\epsilon_d|$), the effective dynamics of the storage mode can be modeled by a two-photon driven-dissipation $\kappa_{2\text{ph}}\cD[\ba^2-\alpha^2]$. Here $\kappa_{2\text{ph}}$ is roughly given by $4|g_{2\text{ph}}|^2/\kappa_d$ and $\alpha$ is given by $\sqrt{-\epsilon_d/g_{2\text{ph}}}$~\cite{Mirrahimi2014}.  The two-photon dissipation rate $\kappa_{2\text{ph}}$  is modulated by the pump power, and the cat amplitude and phase (given by the complex number $\alpha$) are modulated by the resonant drive's amplitude and phase. Strong couplings $g_{2\text{ph}}$, leading to strong 2-photon dissipation rates $\kappa_{2\text{ph}}$, have been achieved in recent experiments~\cite{Leghtas2015,Touzard-PRX-2018}. Remarkably in Ref.~\cite{Touzard-PRX-2018}, the authors engineer a two-photon dissipation rate $\kappa_{2\text{ph}}$ of about 100 times larger than the natural single-photon loss rate and there seems to be room for extra improvements. Finally, the latest experiments~\cite{lescanne-dephasing-2019} have illustrated signatures of the promised exponential bit-flip suppression (with the cat size) for the induced cat qubits. 

One important question to answer is whether the bit-flip suppression induced by the two-photon process is affected by higher order terms in this adiabatic elimination approximation? The answer, fortunately, is not in a significant manner. In order to see this result,  one can note that the states $\{|\pm\alpha\rangle\otimes|0\rangle\}$ are precise steady states of the two-mode system before adiabatic elimination. Indeed, the protection of the coherent states $|\pm\alpha\rangle$ against local shifts is ensured at all orders because of this stability. The rate of convergence to these states (which can be seen as the rate of protection against such excursions in the phase space) can, however, be modified when considering higher-order terms. However, if these corrections to the two-photon dissipation rate are not too large, the associated protection rate remains larger than the diffusion rate induced by local error mechanisms and, therefore, the bit-flip suppression remains valid.

In parallel to these experiments, the measurement operation $\cM_X$, which is equivalent to measuring the photon-number parity of the cat qubits has been performed in Refs.~\cite{Sun2014,Ofek-Petrenko-Nature-2016}. Following the approach presented in Sec.~\ref{sec:physical_gates}, a measurement fidelity of 98.5\% has been achieved in Ref.~\cite{Ofek-Petrenko-Nature-2016}. Finally, quantum Zeno dynamics can be applied to perform bias preserving rotations around the $Z$ axis of the cat qubit. As explained in Sec.~\ref{sec:physical_gates}, it is enough to turn on a weak resonant drive at the frequency of the storage mode $\omega_a$ in the presence of the two-photon driven dissipation. This experiment was performed in Ref.~\cite{Touzard-PRX-2018}.

The realization of the $X$-operation consists in taking the  same approach as the two-photon driven dissipation and simply varying the phase of the resonant drive $\epsilon_d$  between $0$ and $2\pi$ in a time $T$. This  leads to a dissipation operator $\kappa_{2\text{ph}}\cD[\ba^2-\exp(2i\pi t/T)\alpha^2]$ which implements a bias preserving $X$ operation. In order to remove the phase-flip errors induced by the nonadiabaticity of this variation, one can additionally implement a Hamiltonian of the form $-\Delta\ba^\dag\ba$ with $\Delta=\pi/T$. This is simply done by taking the pump at frequency $2\omega_a-\omega_b-2\Delta$ instead of $2\omega_a-\omega_b$ and furthermore detuning the drive $\epsilon_d$ from resonance by value $\Delta$. These are all simple modifications of the experiments in Refs.~\cite{Leghtas2015,Touzard-PRX-2018,lescanne-dephasing-2019} and should be straightforward. 

\begin{figure}[t!]
\includegraphics[width=\linewidth]{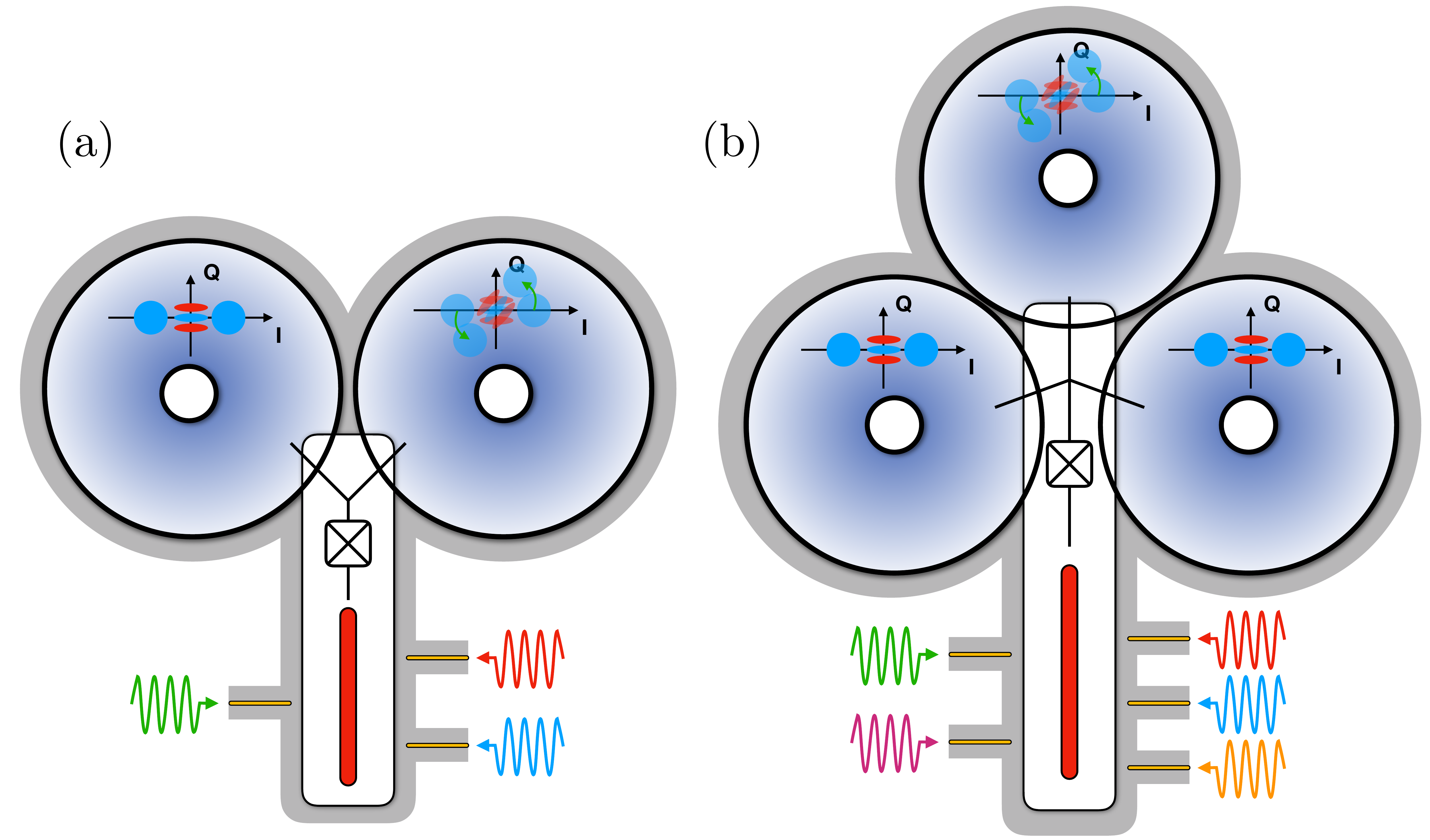}
\caption{\label{fig:CNOT_Toffoli_exp} Proposal for an experimental implementation of bias preserving CNOT and Toffoli gates for cat qubits. (a) Setup for implementing a bias preserving CNOT gate. The cat qubits are encoded in high-Q cylindrical postcavities (in blue, resonance frequencies $\omega_a$ and $\omega_b$). The two cavities are coupled via a Y-shape transmon as in~\cite{Wang-Science-2016} to a low-Q stripline resonator (in red, resonance frequency $\omega_d$) playing the role of the dump mode. The system is driven with three microwave pumps at frequencies $\omega_1=2\omega_b-\omega_d$, $\omega_2=(\omega_a-\omega_d)/2$, $\omega_3=\omega_d$. (b) Similar setup for implementing a bias preserving Toffoli gate with three cat qubits  encoded in high-Q postcavities (frequencies $\omega_a, \omega_b, \omega_c$) all coupled to a single stripline resonator (frequency $\omega_d$). The system is driven with five microwave pumps at frequencies $\omega_1=2\omega_c-\omega_d$, $\omega_2=\omega_a+\omega_b-\omega_d$, $\omega_3=(\omega_a-\omega_d)/2$, $\omega_4=(\omega_b-\omega_d)/2$, and $\omega_5=\omega_d$.}
\end{figure} 

Let us now discuss the realization of the CNOT and Toffoli gates. For the CNOT gate between two cat qubits encoded in storage modes $\ba$ and $\bb$, further to a two-photon driven dissipation modeled by the dissipator $\kappa_{2\text{ph}}\cD[\ba^2-\alpha^2]$, one requires to also implement a time-dependent dissipator given by $\kappa_{2\text{ph}}\cD[\bb^2-\alpha^2-\frac{\alpha}{2}(1-e^{\frac{2i\pi t}{T}})(\ba-\alpha)]$. In order to implement such a dissipation operator, we propose to couple the two storage modes $\ba$ and $\bb$ to a dump mode $\bd$, using a Y-shape transmon similar to~\cite{Wang-Science-2016} (see Fig.~\ref{fig:CNOT_Toffoli_exp}). Driving the dump mode at three different frequencies $\omega_1=2\omega_b-\omega_d$, $\omega_2=(\omega_a-\omega_d)/2$, and $\omega_3=\omega_d$, one can engineer an interaction Hamiltonian of the form
\begin{multline*}
H_{\text{CNOT}}=(g_{bd}\bb^2\bd^\dag+g_{bd}^*\bb^{2\dag}\bd)\\+(g_{ad}\ba\bd^\dag+g_{ad}^*\ba^\dag\bd)+(\epsilon_d\bd^\dag+\epsilon_d^*\bd).
\end{multline*}
In this interaction Hamiltonian, the first term $(g_{bd}\bb^2\bd^\dag+g_{bd}^*\bb^{2\dag}\bd)$ models the exchange of two storage photons at frequency $\omega_b$ with one dump photon at frequency $\omega_d$ via a pump photon at frequency $\omega_1$.  The second term $(g_{ad}\ba\bd^\dag+g_{ad}^*\ba^\dag\bd)$ models the exchange of one storage photon at frequency $\omega_a$ with one dump photon at frequency $\omega_d$ via two pump photons at frequency $\omega_2$. The amplitudes and phases of $g_{bd}$ and $g_{ad}$ are modulated by the amplitude and phase of the corresponding pumps. Finally, the last term $(\epsilon_d\bd^\dag+\epsilon_d^*\bd)$ models the resonant interaction of the drive at frequency $\omega_d$ with the dump mode. Similarly to the driven two-photon dissipation, one can adiabatically eliminate the highly dissipative dump mode to achieve an effective dissipation operator 
$$
\kappa_{2\text{ph}}\cD[\bb^2+c_a\ba+c]
$$
where the dissipation rate $\kappa_{2\text{ph}}$ is roughly given by $4|g_{bd}|^2/\kappa_d$ ($\kappa_d$ being the loss rate of the dump mode), the complex constant $c_a$ is given by $g_{ad}/g_{bd}$ and the complex constant $c$ by $\epsilon_d/g_{bd}$. Similarly to the $X$-operation, it is clear that by varying the amplitudes and phases of the pump at frequency $\omega_2$ and the resonant drive at frequency $\omega_d$, one can engineer a dissipation operator with time-varying constants $c_a$ and $c$ is given, respectively, by
$$
c_a(t)=-\frac{\alpha}{2}\left(1-e^{\frac{2i\pi t}{T}}\right),\qquad c(t)=-\frac{\alpha^2}{2}\left(1+e^{\frac{2i\pi t}{T}}\right).
$$
This result corresponds to the dissipator required for the bias preserving CNOT operation. Importantly, the time-dependent function $c_a$ takes the value 0 at times $t=0$ and $t=T$. For this reason, before and after the gate, the two cat qubits involved in the CNOT are defined by their own local oscillators. The fluctuations of the pumps during the execution of the gate merely result in a slight modification of the geometric paths taken. This modification can lead only to small fluctuations of the geometric phase and therefore an effective phase-flip type error. The phase-flip probability induced by the nonadiabaticity of the evolution can be reduced by adding the effective Hamiltonian $\widetilde H_{\text{CNOT}}= \frac12 \frac \pi T \frac{ \hat a -\alpha}{2\alpha} \otimes (\hat{b}^\dag \hat b-|\alpha|^2) +$ H.c. . Such a Hamiltonian has also been recently implemented using a detuned parametric pumping method~\cite{Touzard-PRL-2019}. 

In order to realize a bias preserving Toffoli gate between three cat qubits encoded in storage modes $\ba$, $\bb$ and $\bc$, further to two-photon driven dissipations modeled by $\kappa_{2\text{ph}}\cD[\ba^2-\alpha^2]$ and $\kappa_{2\text{ph}}\cD[\bb^2-\alpha^2]$, we require to implement a time-dependent dissipator given by $\kappa_{2\text{ph}}\cD[\hat{c}^2 -\alpha^2+\frac{1}{4}(1-e^{\frac{2i\pi t}{T}})(\ba\bb-\alpha(\ba+\bb)+\alpha^2)]$. Similarly to the CNOT gate, we propose to couple the three modes to a highly dissipative dump mode as shown in Fig.~\ref{fig:CNOT_Toffoli_exp}. By driving the dump mode at five different frequencies $\omega_1=2\omega_c-\omega_d$, $\omega_2=\omega_a+\omega_b-\omega_d$, $\omega_3=(\omega_a-\omega_d)/2$, $\omega_4=(\omega_b-\omega_d)/2$, and $\omega_5=\omega_d$, one can engineer an effective interaction Hamiltonian of the form
\begin{multline*}
H_{\text{Toffoli}}=(g_{cd}\bc^2\bd^\dag+g_{cd}^*\bc^{\dag 2}\bd)+(g_{abd}\ba\bb\bd^\dag+g_{abd}^*\ba^{\dag}\bb^\dag \bd)\\
+(g_{ad}\ba\bd^{\dag}+g_{ad}^*\ba^\dag\bd)+(g_{bd}\bb\bd^{\dag}+g_{bd}^*\bb^\dag\bd)+(\epsilon_d\bd^\dag+\epsilon_d^*\bd).
\end{multline*}
Once again, all these effective terms are achieved in a parametric manner and using the 4-wave mixing property of the Josephson junction. The amplitude and phase of each interaction term can be modulated by the amplitude and phase of the associated pump. After the adiabatic elimination of the dump mode, we achieve a dissipation operator 
$$
\kappa_{2\text{ph}}\cD[\bc^2+c_{ab}\ba\bb+c_a\ba+c_b\bb+c],
$$
where $\kappa_{2\text{ph}}$ is given by $4g_{cd}^2/\kappa_d$, and the complex constants $c_{ab}=g_{abd}/g_{cd}$, $c_a=g_{ad}/g_{cd}$, $c_b=g_{bd}/g_{cd}$, $c=\epsilon_d/g_{cd}$. By varying the amplitudes and phases of the pumps in time, we obtain time-varying constants
\begin{align*}
c_{ab}(t)&=\frac{1}{4}\left(1-e^{\frac{2i\pi t}{T}}\right),\quad c(t)=-\frac{\alpha^2}{4}\left(3+e^{\frac{2i\pi t}{T}}\right),\\
c_{a}(t)&=c_b(t)=-\frac{\alpha}{4}\left(1-e^{\frac{2i\pi t}{T}}\right).
\end{align*}
This implements a bias preserving Toffoli gate between the cat qubits encoded in the three modes $\ba,\bb$ and $\bc$. Here again, it should be noted that the functions $c_{ab}$, $c_a$, $c_b$ vanish  at the beginning and at the end of the gate execution, so that  each cat qubit gets back to being defined by its own local oscillators. Similarly to the CNOT gate, the pump fluctuations during the gate result only in a slight increase in the rate of phase-flip type errors, but do not lead to unsuppressed bit-flip-type ones.
In order to reduce the phase-flip probability induced by the nonadiabaticity, we use an additional Hamiltonian $\widetilde H_{\text{Toffoli}}= -\frac12 \frac \pi T \frac{\hat a-\alpha}{2\alpha}\otimes \frac{\hat b-\alpha}{2\alpha}\otimes (\hat{c}^\dag \hat c-|\alpha|^2)+$ H.c.. Such a Hamiltonian can also be implemented in a similar manner to $\widetilde H_{\text{CNOT}}$. A more detailed discussion of such a Hamiltonian synthesis together with the error probability enhancements is out of the scope of the present paper and will be the subject of a forthcoming paper. 

Finally, we note that the bias preserving gate CZ$(\theta)$ is not required  in the fundamental set of bias-preserving operations but may be useful for optimizing the realization of certain quantum algorithms. Such an operation can also be achieved following an approach based on quantum Zeno dynamics, in a similar manner to Z$(\theta)$~\cite{Mirrahimi2014}. The required beam-splitter Hamiltonian $\epsilon_{ZZ}(\ba\bb^\dag+\bb\ba^\dag)$ can be engineered by applying a pump at frequency $(\omega_b-\omega_a)/2$. The exchange of the photons between the two modes is therefore mediated by the 4-wave mixing property of the Josephson junction via two pump photons. 

All the above implementations appear to be rather straightforward extensions of the existing parametric methods, for instance, for two-photon driven dissipation. We expect that the experiments with similar devices and similar parameter regimes as those in the cited references lead to bias preserving gates with phase-flip error probabilities below the threshold of the repetition code. For instance following the error analysis in Sec.~\ref{sec:error_analysis}, we anticipate that a ratio of 1000 between the engineered two-photon decay rate $\kappa_{2\text{ph}}$ and the natural single-photon loss rate $\kappa_{1\text{ph}}$ will put us well below the threshold of the repetition code~\cite{Wang-Preskill-2003}, such that with a cat size of $|\alpha|^2=10$ and with a few tens of repetition modes per logical qubit, we achieve  logical error probabilities of the order $10^{-9}$ for a universal gate set.  This appears to be a huge overhead reduction with respect to state of art quantum error correction approaches. 

\section{Conclusions}\label{sec:conc}
Fault-tolerant computation with protected logical qubits represents a major experimental challenge. The surface code provides a viable solution as it exhibits relatively high accuracy thresholds. However, this property comes at the cost of a tremendous experimental overhead. Indeed, in a realistic implementation of quantum algorithms, the vast majority of operations serve to protect the information rather than performing the computation itself. Furthermore, while certain operations can be performed in a topologically protected manner, certain others (e.g. non-Clifford gates) require magic states  preparation, distillation and injection adding further complexity.  In this paper, we have proposed an alternative approach by replacing the 2D surface code with a 1D repetition code where the physical two-level systems are replaced by cat qubits. Remarkably, we obtain a universal set of topologically protected logical gates with no need for magic states.  The apparent trick lies in the fact that the 2D phase space of the cat qubits are exploited to perform nontrivial operations. In this sense a line of cat qubits has the same properties as an effective 3D system. 

Furthermore, we show that this approach is readily exploitable at an experimental level and requires only minor modifications of previous realizations. A numerical analysis indicates that the parameter regimes close to the ones  achieved in the field of superconducting circuits, result in  effective error probabilities  below the accuracy threshold of the repetition code. Therefore, this scheme is a promising candidate for a first demonstration of a universal set of fully protected logical quantum gates.  Encouragingly, we expect that small logical error rates of $10^{-9}$ could be achieved with a few tens of cat qubits and a mean photon number of about 10. A more thorough error analysis based on a mathematical analysis of the effective error mechanisms and the numerical implementation of an optimal decoding strategy will be subject of a forthcoming paper.

\appendix
\section{A no-go theorem for bias preserving quantum gates}\label{append:nogo}

As we see in the paper, the extra degree freedom associated to the complex amplitude $\alpha$ in the cat qubits enables us to perform a set of nontrivial operations such as the CNOT  or the Toffoli gate in bias-preserving manner. In this appendix, we show the crucial role played by this extra degree of freedom. Indeed, we will prove that such gates cannot be performed in a bias-preserving manner with  two-level systems. The analysis of this appendix should be extendable to the case of qubits encoded in qudits with a finite number of levels. However, as the dimension of the space in which the qubit is encoded increases, the bias could be approximately preserved. 

We will focus on the case of the CNOT gate but a similar analysis can be performed for the Toffoli gate. Throughout this section, we will call $\cU(4)$ the Lie group of unitary operators on two two-level systems and $\csu(4)$ the associated Lie algebra.  We also assume that we are dealing with qubits that are only susceptible to phase-flip errors. We define a bias preserving gate to be a gate that does not transform phase-flip errors to  bit-flip ones. Here is a more precise definition.  
\begin{dfn}
We call a unitary operation $U\in \cU(4)$   bias preserving, if
$$
[U Z_{1,2}U^\dag,Z_{1,2}]=0 .
$$ 
\end{dfn} 
Indeed, the operators $iZ_1$ and $iZ_2$ and similarly $iU Z_{1,2}U^\dag$ are members of the Lie algebra $\csu(4)$ and therefore can be written in the basis of two-qubit Pauli operators. It is therefore easy to see that the above condition is equivalent to saying that $U Z_{1,2}U^\dag$ is a linear combination of the three Pauli operators $Z_1,Z_2,Z_1Z_2$. This e.g. means that 
\begin{align*}
U Z_1 U^\dag&=c_1Z_1+c_2Z_2+c_{12}Z_1Z_2 ~ \Rightarrow \\
 &U Z_1= (c_1Z_1+c_2Z_2+c_{12}Z_1Z_2) U.
\end{align*}
Therefore, a $Z_1$ error before the unitary operation can only lead to $Z_1,Z_2$ and $Z_1Z_2$ errors after the operation. 

We note that, the CNOT  or Toffoli gates are members of this set of bias-preserving operations. We will see however, that they cannot be realized in a bias-preserving manner. We have the following Lemma:
\begin{lem}\label{lem:lie}
The set 
$$
\cB=\{U\in \cU(2)~|~[U^\dag Z_{1,2}U,Z_{1,2}]=0\}
$$
is a Lie subgroup of $\cU(2)$.
\end{lem}
\textit{Proof.} We start by proving that $\cB$ is a group. It clearly includes the identity operator. Also if $U_1,U_2 \in \cB$, we have
$$
U_1Z_1U_1^\dag=c_1 Z_1 + c_2 Z_2 + c_{12}Z_1Z_2.
$$
Therefore,
\begin{align*}
U_2U_1Z_1&U_1^\dag U_2^\dag = c_1 U_2 Z_1 U_2^\dag + c_2 U_2 Z_2U_2^\dag + c_{12}U_2 Z_1Z_2U_2^\dag\\
&=c_1 U_2 Z_1 U_2^\dag + c_2 U_2 Z_2U_2^\dag + c_{12}U_2 Z_1U_2^\dag U_2 Z_2U_2^\dag \\
&=\tilde c_1 Z_1+\tilde c_2 Z_2 +\tilde c_{12} Z_1Z_2.
\end{align*}
Therefore $U_1U_2\in\cB$.  We only need to prove that, if $U\in \cB$ then $U^\dag \in \cB$. We have
\begin{align*}
U Z_1U^\dag&=r_1 Z_1 + r_2 Z_2 +r_{12}Z_1Z_2,\\
U Z_2U^\dag&=s_1 Z_1 + s_2 Z_2 +s_{12}Z_1Z_2.
\end{align*}
We note that $r_1,r_2,r_{12}$ cannot simultaneously vanish (similarly for  $s_1,s_2,s_{12}$). Also, we note that the two vectors $(r_1,r_2,r_{12})$ and $(s_1,s_2,s_{12})$ are necessarily orthogonal. In order to see this, we note that by multiplying the above equations and taking the trace of both sides we get
$$
r_1s_1+r_2s_2+r_{12}s_{12}=0.
$$  
Now we multiply the above equations from left by $U^\dag$ and from right by $U$:
\begin{align*}
Z_1&=r_1 U^\dag Z_1 U+ r_2U^\dag Z_2 U+r_{12}U^\dag Z_1Z_2 U,\\
Z_2&=s_1U^\dag Z_1 U+ s_2U^\dag Z_2 U+s_{12}U^\dag Z_1Z_2 U.
\end{align*}
Furthermore, the product of the above equations give
\begin{multline*}
Z_1Z_2 =(r_2s_{12}+s_2r_{12})Z_1\\
+(r_1s_{12}+s_1r_{12})Z_2+(r_1s_2+s_1r_2)Z_1Z_2.
\end{multline*}
We note that this can not be a linear combination of  $Z_1$ and $Z_2$, which means that the vector $(r_2s_{12}+s_2r_{12},r_1s_{12}+s_1r_{12},r_1s_2+s_1r_2)$ is linearly independent from the orthogonal vectors $(r_1,r_2,r_{12})$ and $(s_1,s_2,s_{12})$. This means the matrix provided by these vectors can be inverted and therefore the operators  $U^\dag Z_1 U$ and $U^\dag Z_2 U$ can also be written as a linear combination of $Z_1,Z_2,Z_1Z_2$. Thus, $U^\dag \in \cB$ and we have therefore shown that $\cB$ is a sub-group of $\cU(4)$.

In order to prove that it is a Lie sub-group, we note that $f(U)=\Big([U^\dag Z_jU,Z_{k}]\Big)_{j,k=1,2}$ is a continuous function of $U$. Furthermore $\cB$ is defined as the pre-image of the set $\{(0,0,0,0)\}$ which is a closed set.  Therefore $\cB$ is topologically closed. A topologically closed sub-group of a Lie group is a Lie sub-group (Cartan's theorem) and therefore the proof is complete. 
$\square$

We now follow ideas that are very similar to the analysis of~\cite{Eastin-Knill-PRL-2009}. The Lie group $\cB$ can be partitioned into cosets of the connected component of the identity that we call $\cC$. $\cC$ is itself a Lie sub-group of $\cB$. This set of cosets is the quotient group $\cB/\cC$. The main result of this appendix can be resumed in the following theorem:
\begin{thm}\label{thm:main}
The unitary operator CNOT is not a member of $\cC$. This means that CNOT cannot be continuously obtained from identity in a bias preserving process.
\end{thm}

\textit{Proof.} As $\cC$ is a connected Lie group, any element $C\in\cC$ can be written as $C=\Pi_k e^{iD_k}$, where $D_k$ is in $\cc$, the Lie algebra of $\cC$. Now, note that for any $\epsilon\in\RR$ and any $D\in \cc$, the operator $e^{i\epsilon D}$ is also in $\cC$ and therefore satisfies 
$$
[e^{i\epsilon D}Z_{1,2}e^{-i\epsilon D},Z_{1,2}]=0.
$$
Taking the derivative with respect to $\epsilon$ at $\epsilon=0$, we get
$$
[[D,Z_j],Z_k]=0,\qquad j,k=1,2.
$$
Noting that $D$ is necessarily a linear combination to two-qubit Pauli operators, it is the same for $[D,Z_j]$ and therefore
\begin{align*}
[D,Z_1]&=r_1Z_1+r_2Z_2+r_{12}Z_1Z_2,\\
[D,Z_2]&=s_1Z_1+s_2Z_2+s_{12}Z_1Z_2.
\end{align*}
The only possibility for such a combination is that all the coefficients vanish. Therefore $[D,Z_{1,2}]=0$, or equivalently
$$
D=c_0 I+c_1Z_1+c_2Z_2+c_{12}Z_1Z_2.
$$
Therefore, the Lie algebra $\cc$ is spanned by $I, Z_1, Z_2, Z_1Z_2$, which means that the associated Lie group does not include CNOT.
$\square$

\begin{acknowledgments}
We thank Shruti Puri for many enlightening discussions. The no-go theorem in the appendix was proven as a followup to a discussion  with Shruti Puri, Steve Flammia, Steven Girvin and Liang Jiang. We thank them for their helpful comments. We also thank Yale Quantum Institute for hosting us while we were working on this paper. This work has been supported by Region Ile-de-France in the framework of DIM SIRTEQ. We also acknowledge financial support by ARO under Grant No. W911NF-18-1-0212. 
\end{acknowledgments}

\end{document}